\documentclass[12pt]{article}
\usepackage{amsmath,epsfig,euscript,url}

\def\vslash{\rlap{\hspace{0.02cm}/}{v}}

\usepackage{bm}

\begin{document}

\begin{titlepage}

\begin{flushright}
WSU-HEP-1701\\
February 28, 2017\\
\end{flushright}

\vspace{0.7cm}
\begin{center}
\Large\bf\boldmath
On HQET and NRQCD Operators of Dimension 8 and Above
\unboldmath
\end{center}

\vspace{0.8cm}
\begin{center}
{\sc Ayesh Gunawardana and Gil Paz}\\
\vspace{0.4cm}
{\it 
Department of Physics and Astronomy \\
Wayne State University, Detroit, Michigan 48201, USA 
}

\end{center}
\vspace{1.0cm}
\begin{abstract}
  \vspace{0.2cm}
  \noindent
Effective field  theories such as Heavy Quark Effective Theory (HQET) and Non Relativistic Quantum Chromo-(Electro-) dynamics NRQCD (NRQED) are indispensable tools in controlling the effects of the strong interaction.  The increasing experimental precision requires the knowledge of higher dimensional operators.  We present a general method that allows for an easy construction of HQET or NRQCD (NRQED) operators that contain two heavy quark or non-relativistic fields and any number of covariant derivatives. As an application of our method, we list these terms in the $1/M^4$ NRQCD Lagrangian, where $M$ is the mass of of the spin-half field.  
  
\end{abstract}
\vfil

\end{titlepage}

\section{Introduction}
Effective field  theories such as Heavy Quark Effective Theory (HQET) and Non Relativistic Quantum Chromo-(Electro-) dynamics NRQCD (NRQED) are indispensable tools in controlling the effects of the strong interaction in diverse areas such as flavor physics, see e.g  \cite{Neubert:1993mb, Manohar:2000dt, Petrov:2016azi}, and proton structure effects in hydrogen-like systems, see e.g. \cite{Pineda:2002as,Pineda:2004mx, Nevado:2007dd, Hill:2011wy, Birse:2012eb, Hill:2012rh, Gorchtein:2013yga, Alarcon:2013cba, Peset:2014yha, Peset:2014jxa, Dye:2016uep}. The increasing experimental precision requires the knowledge of higher dimensional operators. For example, in extracting $|V_{cb}|$ from inclusive $B$ decays, matrix elements of HQET operators of dimension seven and eight  are now used \cite{Gambino:2016jkc}. These operators contain two heavy quarks fields and four or five covariant derivatives.  As was shown in \cite{Mannel:2010wj}, see also \cite{Balk:1993ev, Dassinger:2006md}, at dimension seven, four spin-independent and five spin-dependent matrix elements are needed. At dimension eight, seven spin-independent and eleven spin-dependent matrix elements are needed . A natural question arises: are these \emph{all} the possible operators at each dimension? In particular it could be that semileptonic decays considered in \cite{Mannel:2010wj} depend only on a subset of the possible HQET operators.

The HQET  and NRQCD Lagrangians up to and including power $1/M^3$,  where $M$ is the mass of of the spin-half field, were given in \cite{Manohar:1997qy}. These include operators of dimension seven and below. Although these two theories differ in the their power counting, the two Lagrangians can be related via field redefinitions. The dimension-seven Lagrangian contains six spin-independent operators and five spin-dependent operators. For the spin-dependent operators the number is  the same as the numbers of the  spin-dependent matrix elements considered in \cite{Mannel:2010wj}, while the spin independent  number of operators is different. Why is there a difference and what is the relation between these two bases? 

More recently, the NRQED Lagrangian up to and including power $1/M^4$ was calculated in \cite{Hill:2012rh}. It includes NRQED operators of dimension eight and below.  The Lagrangian was constructed by considering all the possible rotationally invariant, $P$ and $T$ even, Hermitian combinations of $iD_t$, $i\bm{D}$, $\bm{E}$, $\bm{B}$, and $\bm{\sigma}$. The analogous construction of the NRQED Lagrangian up to $1/M^2$ was explicitly demonstrated in \cite{Paz:2015uga}. For higher power of $1/M$, corresponding to higher dimensional operators, this construction becomes  tedious. There can be different choices for the form of the operators. It is not immediately clear if a pair of operators is linearly independent and what is the total number of linearly independent operators.  It would be useful to find a simpler way to construct these operators. Furthermore, the $1/M^4$ NRQED Lagrangian contains four spin-independent and eight spin-dependent operators. This is less than the number of matrix elements considered in \cite{Mannel:2010wj}.  Presumably the rest correspond to NRQCD operators that do not exist for NRQED. What are they? 

In this paper we address  all of these questions. We show that by considering the general decomposition of the diagonal matrix elements of pseudo-scalar heavy meson $H$ of the form $\langle H|\bar h\, iD^{\mu_1}\dots\, iD^{\mu_n}(s^\lambda)h|H\rangle$, where $iD^\mu$ is the covariant derivative and $s^\lambda$ is the four-dimensional generalization of the Pauli matrices, into linearly independent tensors, we can determine the form and number of linearly independent operators at each dimension (up to some possible color factors). This method was used in \cite{Mannel:1994kv} for dimension five and six HQET operators. We improve on \cite{Mannel:1994kv}, by applying constraints from various symmetries and generalize it to an arbitrary operator dimension\footnote{While this work was in progress, \cite{Heinonen:2016cwm} appeared. The appendix of  \cite{Heinonen:2016cwm} lists a tensor decomposition for HQET operators up to dimension eight and relates it to \cite{Mannel:2010wj}.}. We will show that one can make one-to-one correspondence between these HQET operators and NRQCD (NRQED) operators. This allows in principle to construct the bilinear NRQCD (NRQED) Lagrangian at any given order in $1/M$.   

We present several applications of our method. We relate the Manohar \cite{Manohar:1997qy} and Mannel-Turczyk-Uraltsev \cite{Mannel:2010wj} bases for the dimension-seven operators.  We relate the dimension-eight operators of  \cite{Mannel:2010wj} to the NRQED operators of  \cite{Hill:2012rh}. We analyze the general dimension-nine spin-independent HQET matrix element, not considered so far in the literature, and calculate moments of the leading power shape function up to and including dimension nine HQET operators. Finally, we will present  the bilinear NRQCD Lagrangian at order $1/M^4$.

Throughout this paper we will not discuss four-fermion operators. The main reason is that the one heavy fermion sector can be considered as interacting with another heavy fermion leading to, e.g. higher power NRQCD and NRQED four-fermion operators, see \cite{Bodwin:1994jh, Brambilla:2006ph, Brambilla:2008zg, Hill:2012rh}, or as interacting with a relativistic fermion leading to the four-fermion operators QED-NRQED operators, see \cite{Hill:2012rh, Dye:2016uep}. Each case should be considered separately depending on the application and is beyond the scope of this paper.  Similarly we will not consider pure gauge operators. 

The rest of the paper is structured as follows. We present our notation, the general method, and a tabulation of the operators up to and including dimension eight  in section \ref{General}. We compare our basis to  the known dimension seven and eight HQET operators in section \ref{HQET}. We compare  our basis to  the known NRQED and NRQCD operators in section \ref{NRQED}.  In section \ref{Applications} we analyze the general dimension-nine spin-independent HQET matrix element, calculate moments of the leading power shape function, and give the $1/M^4$ NRQCD Lagrangian. We present our conclusions in section \ref{Conclusions}. 

After the first version of this paper appeared on arxiv.org, a paper by Kobach and Pal appeared on arxiv.org that uses a Hilbert series to construct an operator basis for NRQED and NRQCD/HQET up to and including dimension eight \cite{Kobach:2017xkw}. As the authors of \cite{Kobach:2017xkw} explain ``While the Hilbert series can count the number of operators that are invariant under the given symmetries, it does not say how the indices within each operator are contracted. In general, this needs to be done by hand." While analyzing the color indices they have pointed out some operators with the same Lorentz structure but different color structures that were omitted in the first version of this paper. We address the issue of possible multiple color structures and correct this omission in section \ref{General.1}.

\section{General Method} \label{General}
\subsection{General considerations} \label{General.1}  
We begin by presenting the notation we use. We define the metric as $g^{\mu\nu}=$diag$(1,-1,-1,-1)$. The four velocity $v$ is defined by, $v=p/M$, where $p$ is the particle momentum and $M$ its mass.  We often take $v=(1,0,0,0)$ to simplify the discussion. We follow the standard HQET notation reviewed in \cite{Neubert:1993mb, Manohar:2000dt, Petrov:2016azi}. In particular we denote the heavy quark field as $h$. We will consider diagonal matrix elements of HQET operators between pseudo-scalar meson states $H$ containing a heavy quark.  Between heavy quark fields the Dirac basis reduces to four matrices \cite{Mannel:1994kv}. Following \cite{Mannel:1994kv} we take the basis to be $\left\{1, s^\lambda\right\}$, where the spin matrices $s^\lambda$ are a generalization of the Pauli matrices for a frame moving with velocity $v$. In particular, $(1+\vslash)\gamma^\lambda\gamma^5(1+\vslash)\equiv 4s^\lambda$.
The matrices $s^\lambda$ are orthogonal to $v$, i.e. $v\cdot s=0$. We follow \cite{Manohar:1997qy} and define the covariant derivative as $D^\mu=\partial^\mu+igA^{\mu a}T^a=\left(D^0,-\bm{D}\right)$, where $D^0=\partial/\partial t+igA^{0 a}T^a$ and $\bm{D}=\bm{\nabla}-ig\bm{A}^{a}T^a$. The (chromo-) electric and magnetic fields are defined as $\bm{E}=(-i/g)[D^0,\bm D]$ and $\bm{B}^i=\epsilon^{ijk}(i/2g)[\bm{D}^j,\bm{D}^k]$. For the case of HQET and NRQCD $\bm{E}\equiv \bm{E}_aT^a$ and $\bm{B}\equiv \bm{B}_aT^a$, where $T^a$ is an $SU(3)$ color matrix.  As usual, $[X,Y]\equiv XY-YX$, $\{X,Y\} \equiv XY+YX$ denote commutators and anti-commutators.

We analyze matrix element of the form $\langle H|\bar h\, iD^{\mu_1}\dots\, iD^{\mu_n}(s^\lambda)h|H\rangle$, where $n$ is a positive integer. We would like  to collect the constraints on them as a result of heavy quark symmetry, parity, time-reversal, and Hermitian conjugation before the tensor decomposition.   

The equation of motion of HQET imply that $iv\cdot Dh=0$. As a result, multiplying the matrix element by $v_{\mu_n}$ gives zero\footnote{More accurately, the $1/M$ corrections to this equation give rise to higher dimensional operators, see e.g.  \cite{Petrov:2016azi}. One can therefore impose this equation order by order in the $1/M$ expansion.}. Since this is a forward matrix element, we find that if we multiply the matrix element by $v_{\mu_1}$ we will get zero too. Therefore HQET implies that $\mu_1$ and $\mu_n$ must be orthogonal to $v$ \cite {Mannel:1994kv}.  An analogous relation holds for the NRQED or NRQCD operators. By a field redefinition one can always eliminate operators of the form $\dots\, iv\cdot D\psi$ or $\psi^\dagger iv\cdot D\dots\, $ from the Lagrangian. See \cite{Paz:2015uga} for an explicit example. This property allows us to treat NRQCD operators similar to HQET operators. 

The effective field theories we consider are invariant under parity and time-reversal. As a result the matrix elements have definite transformation properties under these symmetries. Consider $v$ first. Under parity the meson four-momentum changes as $p=(p^0,\vec p\,)\stackrel{P}{\to}p_P\equiv(p^0,-\vec p\,)$. As a result $v\stackrel{P}{\to} (v^0,-\vec v\,)$. Under time-reversal the meson four momentum changes as $p=(p^0,\vec p\,)\stackrel{T}{\to}p_T\equiv(p^0,-\vec p\,)$. As a result $v\stackrel{T}{\to} (v^0,-\vec v\,)$. Notice that in both cases for the standard choice of $v=(1,0,0,0)$, $v$ does not change. Under the combined operation of parity and time-reversal $p=(p^0,\vec p\,)\stackrel{PT}{\to}p_{PT}\equiv(p^0,\vec p\,)$. As a result $v\stackrel{PT}{\to} (v^0,\vec v\,)$, i.e. any choice of $v$ is invariant under $PT$. Consider next the covariant derivative  $iD^\mu$. Under parity $iD^\mu\stackrel{P}{\to}(-1)^\mu iD^\mu$, where $(-1)^\mu=1$ for $\mu=0$ and $(-1)^\mu=-1$ for $\mu=1,2,3$. Due to presence of $i$, under time-reversal $iD^\mu\stackrel{T}{\to}(-1)^\mu iD^\mu$. Under the combined operation of parity and time-reversal $iD^\mu\stackrel{PT}{\to} iD^\mu$. Finally we need to consider the transformation of $\bar h h$ and $\bar h s^\lambda h$. If $\psi$ is the full QCD quark field, $\bar h h=\bar \psi\left(1+\vslash\right) \psi/2$ and  $\bar h s^\lambda h=\bar \psi(1+\vslash)\gamma^\lambda\gamma^5(1+\vslash)\psi/4$. From the known transformation properties of the Dirac bilinear we find that $\bar h h$ is invariant under parity, time-reversal, and the combined operation of parity and time-reversal.   Similarly $\bar h s^\lambda h$ transforms as $\bar h s^\lambda h\stackrel{P}{\to} -(-1)^\lambda\,\bar h s^\lambda h$ under parity, $\bar h s^\lambda h\stackrel{T}{\to} (-1)^\lambda\,\bar h s^\lambda h$  under time-reversal and  $\bar h s^\lambda h\stackrel{PT}{\to} -\,\bar h s^\lambda h$ under the combined operation of parity and time-reversal. Combining all of these allows us to show that under the combined operation of parity and time-reversal, 
\begin{eqnarray}
\langle H|\bar h\, iD^{\mu_1}\dots\, iD^{\mu_n}h|H\rangle\stackrel{PT}{=}
\langle H|\bar h\, iD^{\mu_1}\dots\, iD^{\mu_n}h|H\rangle^*\, ,\nonumber\\
\langle H|\bar h\, iD^{\mu_1}\dots\, iD^{\mu_n}s^\lambda h|H\rangle\stackrel{PT}{=}
-\langle H|\bar h\, iD^{\mu_1}\dots\, iD^{\mu_n}s^\lambda h|H\rangle^*,
\end{eqnarray}
where the complex conjugation arises from the anti-linear $T$.
As a result, the spin-independent matrix elements are real, while the spin-dependent matrix elements are imaginary. Constraints from parity are more transparent for the standard choice of $v=(1,0,0,0)$, where $v$ does not change under parity. For this case, $\bar h h$, $\bar h s^\lambda h$, $iD^0$ are even and $i\bm{D}^i$ is odd. As a result, regardless of the spin structure, the matrix elements vanish if they have an odd number of spacelike covariant derivatives.

Hermitian conjugation also restricts the number of linearly independent matrix elements. Since $\bar h h$, $\bar h s^\lambda h$, and $iD^{\mu_i}$ are hermitian, we find that 
\begin{equation}
\langle H|\bar h\, iD^{\mu_1}\dots\, iD^{\mu_n}(s^\lambda)h|H\rangle=\langle H|\left(\bar h\, iD^{\mu_1}\dots\, iD^{\mu_n}(s^\lambda)h\right)^\dagger|H\rangle^*=\langle H |\bar h\, iD^{\mu_n}\dots\, iD^{\mu_1}(s^\lambda)h|H\rangle^*.
\end{equation}
Combining this with the $PT$ constraints implies that the spin-independent (spin-dependent) matrix elements are symmetric (anti-symmetric) under the inversion of the indices. In the following we refer to it as ``inversion symmetry".

Since $H$ is a pseudo-scalar, the matrix element of $\langle H|\bar h\, iD^{\mu_1}\dots\, iD^{\mu_n}(s^\lambda)h|H\rangle$ can only depend on $v_{\mu_i}$ and $g^{\mu_i\mu_j}$ and $\epsilon^{\alpha\beta\rho\sigma}$. Alternatively, we can follow \cite{Mannel:2010wj} and define $\Pi^{\mu\nu}=g^{\mu\nu}-v^\mu v^\nu$. In general $v_\mu\Pi^{\mu\nu}=0$ and $v_\nu\Pi^{\mu\nu}=0$.  For the standard choice of $v=(1,0,0,0)$, $\Pi^{00}=0$ and $\Pi^{ij}=-\delta^{ij}$. Also, since the indices in $\epsilon^{\alpha\beta\rho\sigma}$ cannot all be orthogonal to  $v$, we can replace  $\epsilon^{\alpha\beta\rho\sigma}$ by $\epsilon^{\alpha\beta\rho\sigma}v_\alpha$ without loss of generality. 

Another constraint to keep in mind is that in four dimensions one can have only four independent directions. As a result, certain tensors with more than four indices are not independent. For example, in a  tensor of the form $\Pi^{\mu\nu}\epsilon^{\alpha\beta\rho\sigma}v_\alpha$, relevant for the matrix elements of dimension seven spin-dependent operators, three indices are the same and not all the tensors obtained by permuting indices between $\Pi^{\mu\nu}$ and $\epsilon^{\alpha\beta\rho\sigma}v_\alpha$ are linearly independent. For spin-independent operators a similar constraint arises only starting at dimension eleven operators where the structure $\Pi^{\mu_1\mu_2}\Pi^{\mu_3\mu_4}\Pi^{\mu_5\mu_6}\Pi^{\mu_7\mu_8}$ arises. 

The decomposition gives a correspondence between the operators  $\bar h\, iD^{\mu_1}\dots\, iD^{\mu_n}(s^\lambda)h$ and non-perturbative parameters. Questions such as the linear independence of a given set of operators, and the number of linearly independent operators of a given dimension are answered by considering the vector space of  non-perturbative parameters of a given dimension\footnote{A potential caveat to this argument is that one can imagine an operator that has a zero matrix element. The only such example is the operator $\bar h\, iv\cdot Dh$, which is the first term in the HQET and NRQCD (NRQED) Lagrangians. This term is unique in the sense that it is the only one that includes $iv\cdot D$ in the HQET Lagrangian or $iD_t$ (not in a commutator) in the NRQCD (NRQED) Lagrangian.}. 

Another issue we need to address is that of possible color factors. The covariant derivative $D^\mu=\partial^\mu+igA^{\mu a}T^a$  combines a unit matrix in color space, i.e. a color singlet, and a product of an octet vector field $A^{\mu a}$ and an octet of $SU(3)$ color matrices. By gauge invariance the two must appear together, and the covariant derivative does not have an independent color singlet and color octet parts. Operators constructed from two covariant derivatives can be expressed in terms of a commutator or an anti-commutator of two covariant derivatives. A commutator has only an octet part while the anti-commutator has again both singlet and octet parts that cannot be separated. The case of three covariant derivatives is analogous to that of two covariant derivatives, see section \ref{NRQED}. 

When we consider four covariant derivatives the situation changes. We can now have a product of two commutators of covariant derivatives. Consider for example the NRQCD operator $\psi^\dagger E^i_aT^a  E^i_bT^b \psi$ \cite{Kobach:2017xkw}. It contains the symmetric product of two different $SU(3)$ colors matrices: $\left\{T^a,T^b\right\}=\frac13\delta^{ab}+ d^{abc}T^c$. Now the singlet and octet parts are not connected by gauge invariance and they give rise to two operators with different color structure. Instead of a singlet and an octet we can choose the basis of $\left\{T^a,T^b\right\}$ and $\delta^{ab}$. Thus we have two different operators with two chromo-electric fields:  $\psi^\dagger E^i_a E^i_b\left\{T^a,T^b\right\} \psi$ and $\psi^\dagger E^i_a E^i_b \delta^{ab}\psi$. Only the first one is generated by commutator and anti-commutators of covariant derivatives. The second operator is  generated when we consider the one-loop self-energy corrections to the first operators. Thus a one gluon exchange between $\psi^\dagger$ and $\psi$ in $\psi^\dagger E^i_a E^i_b\left\{T^a,T^b\right\} \psi$ gives the color structures
\begin{equation}
T^c_{ij}\left\{T^a,T^b\right\}_{jk}T^c_{kl}=\left\{T^a,T^b\right\}_{jk}\left(\frac12\delta_{il}\delta_{kj}-\frac16\delta_{ij}\delta_{kl}\right)=\dfrac12\delta^{ab}\delta_{il}-\frac16\left\{T^a,T^b\right\}_{il}
\end{equation}
where $i,j,k,l=1,2,3$ and $a,b=1,...,8$ and we have used a color identity for $T^c_{ij}T^c_{kl}$. In other words, when calculating observables at tree level only $\psi^\dagger E^i_a E^i_b\left\{T^a,T^b\right\} \psi$ appears \cite{Manohar:1997qy}. At one loop we need to consider also $\psi^\dagger E^i_a E^i_b \delta^{ab}\psi$. The case of five covariant derivatives is discussed in sections \ref{sec_SI8} and \ref{sec_SD8}.

In applications to power corrections to inclusive $B$ decays, only the contribution of the dimension five  operators are known with ${\cal O}(\alpha_s)$ Wilson coefficients \cite{Becher:2007tk,Ewerth:2009yr, Alberti:2013}. The dimension six and seven operators are known only with ${\cal O}(\alpha_s^0)$ Wilson coefficients. This explains why the dimension seven operator $\psi^\dagger E^i_a E^i_b \delta^{ab}\psi$ and similar dimension eight operators were not included in \cite{Mannel:2010wj} as was recently pointed out in \cite{Kobach:2017xkw}. The analysis we perform below is sensitive only to the possible Lorentz structure of $\langle H|\bar h\, iD^{\mu_1}\dots\, iD^{\mu_n}(s^\lambda)h|H\rangle$ and it does not distinguish operators that contain $\left\{T^a,T^b\right\}$ from $\delta^{ab}$. These need to be put ``by hand". While not ideal, the main complication arises from the Lorentz indices and it is fairly easy to identify the different colors structures, at least below dimension nine.  We show below how one can identify and enumerate the number of operators that have more than one color structure. Using the constraints discussed above we can now perform the tensor decomposition.

\subsection{Tabulation of spin-independent operators up to dimension eight}
These operators are of the form $\langle H|\bar h\, iD^{\mu_1}\dots\, iD^{\mu_n}h|H\rangle$, where $n=\mbox{operator dimension}-3$.  We decompose the matrix elements of such operators in terms of non-perturbative parameters multiplying the possible tensors allowed by the symmetries. 
\subsubsection{Dimension Three}
For dimension three there are no covariant derivatives and one has, see e.g. \cite{Manohar:2000dt},  
\begin{equation}
\dfrac1{2M_H}\langle H|\bar hh|H\rangle=1.
\end{equation}
\subsubsection{Dimension Four}
Since we have only one covariant derivative $iD^{\mu_1}$, the matrix element must be proportional to $v^{\mu_1}$. Since $iv\cdot Dh=0$ the matrix element must vanish. Thus   
\begin{equation}
\dfrac1{2M_H}\langle H |\bar h\, iD^{\mu_1}h|H\rangle=0.
\end{equation}
\subsubsection{Dimension Five}
Since $iv\cdot Dh=0$ and  $\bar h\,iv\cdot D=0$, the matrix element can only depend on $\Pi^{\mu_1\mu_2}$ and we have 
\begin{equation}\label{SI5} 
\dfrac1{2M_H}\langle H |\bar h\, iD^{\mu_1}iD^{\mu_2}h|H\rangle=a^{(5)}\Pi^{\mu_1\mu_2},
\end{equation}
where $a^{(5)}$ is a non-perturbative parameter. The dimension of  the operator appears in the superscript. Notice that $\Pi^{\mu_1\mu_2}=\Pi^{\mu_2\mu_1}$ as required by the inversion symmetry. 
\subsubsection{Dimension Six}
We need to consider $\langle H |\bar h\, iD^{\mu_1}iD^{\mu_2}iD^{\mu_3}h|H\rangle$. The tensor $\epsilon^{\rho\mu_1\mu_2\mu_3} v_\rho$ is ruled out by parity. This is most easily seen by taking $v=(1,0,0,0)$ which requires $\mu_1,\mu_2,\mu_3$ to be space-like. Hence the matrix element has a an odd number of space-like covariant derivatives and  is zero by parity.  The only possible tensor combination is a product of a $v$ and $\Pi$. We must use $\Pi^{\mu_1\mu_3}$ and we find only one possible non-perturbative parameter:
\begin{equation}\label{SI6} 
\dfrac1{2M_H}\langle H |\bar h\, iD^{\mu_1}iD^{\mu_2}iD^{\mu_3}h|H\rangle=a^{(6)}\Pi^{\mu_1\mu_3}v^{\mu_2}.
\end{equation}
Under inversion $\Pi^{\mu_1\mu_3}v^{\mu_2}\to\Pi^{\mu_3\mu_1}v^{\mu_2}=\Pi^{\mu_1\mu_3}v^{\mu_2}$.
\subsubsection{Dimension Seven}
Here we need more than one tensor structure. We can have a product of two $\Pi$'s or a product of $\Pi$ and two $v$'s. For products of two $\Pi$'s  we can contract $\mu_1$ with $\mu_2,\mu_3,$ or $\mu_4$ using $\Pi$. The other two indices are also contracted  by $\Pi$. In total we have three such combinations of two $\Pi$'s. Using two $v$'s, they can only be contracted  with $\mu_2$ and $\mu_3$ giving us a fourth tensor. In total we have 
\begin{eqnarray}\label{SI7}
\dfrac1{2M_H}\langle H |\bar h\, iD^{\mu_1}iD^{\mu_2}iD^{\mu_3}iD^{\mu_4}h|H\rangle&=&a_{12}^{(7)}\Pi^{\mu_1\mu_2}\Pi^{\mu_3\mu_4}+a_{13}^{(7)}\Pi^{\mu_1\mu_3}\Pi^{\mu_2\mu_4}+\nonumber\\
&+&a_{14}^{(7)}\Pi^{\mu_1\mu_4}\Pi^{\mu_2\mu_3}+b^{(7)}\Pi^{\mu_1\mu_4}v^{\mu_2} v^{\mu_3}. 
\end{eqnarray}
It is easy to check that each tensor separately is invariant under inversion. Our notation for the parameters is such that the subscript denotes the first two indices that are contracted via $\Pi$'s in numerical order, and the dimension of the operators appears in the superscript. We also use a different letters for tensors with a different number of $v$'s.

As was mentioned in the introduction, the NRQED Lagrangian has four spin-independent operators. We will show in section \ref {NRQED} that these can be related to the four operators above. It should be clear already  though that it is easier to tabulate the operators as was done here than to construct them from $\bm{E,D}$, and $\bm B$. 

As was pointed out in \cite{Kobach:2017xkw} and discussed in section \ref{General.1},  there can be more than one color structure for operators constructed from four covariant derivatives. This is most easily seen when one constructs NRQCD operators and then consider the possible color structure, as we do in section \ref{NRQED}. But we can anticipate the result by considering structures of the form $\bar h\left\{[iD^{\mu_i},iD^{\mu_j}],[iD^{\mu_k},iD^{\mu_l}]\right\}h$. It is a symmetric product of two $SU(3)$ color matrices that give rise to two possible color structures: a singlet and an octet. There can be three different structures $\bar h\left\{[iD^{\mu_1},iD^{\mu_2}],[iD^{\mu_3},iD^{\mu_4}]\right\}h$, $\bar h\left\{[iD^{\mu_1},iD^{\mu_3}],[iD^{\mu_2},iD^{\mu_4}]\right\}h$, and $\bar h\left\{[iD^{\mu_1},iD^{\mu_4}],[iD^{\mu_2},iD^{\mu_3}]\right\}h$, corresponding to the possible partitions of four indices into two pairs. 
In order to form scalar operators, we need to multiply these structures by one of the four possible tensors on the right hand side of (\ref{SI7}): $\Pi^{\mu_1\mu_2}\Pi^{\mu_3\mu_4}$, $\Pi^{\mu_1\mu_3}\Pi^{\mu_2\mu_4}$, $\Pi^{\mu_1\mu_4}\Pi^{\mu_2\mu_3}$, and $\Pi^{\mu_1\mu_4}v^{\mu_2} v^{\mu_3}$. We find only two linearly independent combinations from all of the contractions, namely, $a_{13}^{(7)} - a_{14}^{(7)}$, and $b^{(7)}$. We confirm this result in section \ref{NRQCD_SI7}. We conclude that we can form only two such operators with two possible color structures each. Including the possible color structures, there are in total six possible NRQCD (HQET) operators.

\subsubsection{Dimension Eight}\label{sec_SI8}
We have five covariant derivatives, so we must have an odd number of $v$'s. We cannot have five $v$'s and there is only one tensor with 3 $v$'s: $\Pi^{\mu_1\mu_5}v^{\mu_2} v^{\mu_3}v^{\mu_4}$. As a result of the inversion symmetry, tensors with one $v$ must be of the form $v^{\mu_2}\Pi\,\Pi+v^{\mu_4}\Pi\,\Pi$ or  $v^{\mu_3}\Pi\,\Pi$.   All together we find seven possible tensors:
\begin{eqnarray} \label{SI8}
&&\dfrac1{2M_H}\langle H |\bar h\, iD^{\mu_1}iD^{\mu_2}iD^{\mu_3}iD^{\mu_4}iD^{\mu_5}h|H\rangle=a_{12}^{(8)}\left(\Pi^{\mu_1\mu_2}\Pi^{\mu_3\mu_5}v^{\mu_4}+\Pi^{\mu_1\mu_3}\Pi^{\mu_4\mu_5}v^{\mu_2}\right)+\nonumber\\
&&a_{13}^{(8)}\left(\Pi^{\mu_1\mu_3}\Pi^{\mu_2\mu_5}v^{\mu_4}+\Pi^{\mu_3\mu_5}\Pi^{\mu_1\mu_4}v^{\mu_2}\right)+a_{15}^{(8)}\left(\Pi^{\mu_1\mu_5}\Pi^{\mu_3\mu_4}v^{\mu_2}+\Pi^{\mu_1\mu_5}\Pi^{\mu_2\mu_3}v^{\mu_4}\right)+\nonumber\\
&&b_{12}^{(8)}\Pi^{\mu_1\mu_2}\Pi^{\mu_4\mu_5}v^{\mu_3}+b_{14}^{(8)}\Pi^{\mu_1\mu_4}\Pi^{\mu_2\mu_5}v^{\mu_3}+b_{15}^{(8)}\Pi^{\mu_1\mu_5}\Pi^{\mu_2\mu_4}v^{\mu_3}+\nonumber\\
&&c^{(8)}\Pi^{\mu_1\mu_5}v^{\mu_2}v^{\mu_3}v^{\mu_4}.
\end{eqnarray}
Our notation is as above, but we use different letters for the $v^{\mu_2}\Pi\,\Pi+v^{\mu_4}\Pi\,\Pi$ and $v^{\mu_3}\Pi\,\Pi$ tensors. 

We also need to consider the issue of possible color structures.  Multiple colors structures for a given operator arise from the anti-commutator of two color octets. For five covariant derivatives there are two possibilities of color octets: $[iD^{\mu_i},iD^{\mu_j}]$ and $[iD^{\mu_k},[iD^{\mu_l},iD^{\mu_m}]]$. If we combine them together we get two structures\footnote{A third possible structure $\bar h\left\{[iD^{\mu_i},iD^{\mu_j}],[iD^{\mu_l},[iD^{\mu_m},iD^{\mu_k}]]\right\}h$ is related to the first two by the Jacobi identity.}  $\bar h\left\{[iD^{\mu_i},iD^{\mu_j}],[iD^{\mu_k},[iD^{\mu_l},iD^{\mu_m}]]\right\}h$ and $\bar h\left\{[iD^{\mu_i},iD^{\mu_j}],[iD^{\mu_m},[iD^{\mu_k},iD^{\mu_l}]]\right\}h$. There are $ {5 \choose 2}\times 2=20$ such structures. We can also combine $\left\{[iD^{\mu_i},iD^{\mu_j}],[iD^{\mu_k},iD^{\mu_l}]\right\}$ with an anti-commutator of a fifth covariant derivative\footnote{using a commutators does not give a new structures since $[iD^{\mu_m},\left\{[iD^{\mu_i},iD^{\mu_j}],[iD^{\mu_k},iD^{\mu_l}]\right\}]=\left\{[iD^{\mu_i},iD^{\mu_j}],[iD^{\mu_m},[iD^{\mu_k},iD^{\mu_l}]]\right\}+\left\{[iD^{\mu_k},iD^{\mu_l}],[iD^{\mu_m},[iD^{\mu_i},iD^{\mu_j}]]\right\}$.}: $\bar h\left\{iD^{\mu_m},\left\{[iD^{\mu_i},iD^{\mu_j}],[iD^{\mu_k},iD^{\mu_l}]\right\}\right\}h$. There are $ {5 \choose 1}\times 3=15$ such structures.  Contracting each of the possible structure with the tensors on the left hand side of (\ref{SI8}), we find  only one non-zero linear combination: $a^{(8)}_{12} - a^{(8)}_{15} - b^{(8)}_{14} + b^{(8)}_{15}$ from $\bar h\left\{iD^{\mu_m},\left\{[iD^{\mu_i},iD^{\mu_j}],[iD^{\mu_k},iD^{\mu_l}]\right\}\right\}h$. We will obtain the same result in section \ref{NRQCD_SI8}. Including the two possible color structures there are eight operators in total.

\subsection{Tabulation of spin-dependent operators up to dimension eight}
These operators are of the form $\langle H |\bar h\, iD^{\mu_1}\dots\, iD^{\mu_n}s^\lambda h|H\rangle$, where $n=\mbox{operator dimension}-3$. Their matrix elements are complex. We decompose the matrix elements in terms of non-perturbative parameters multiplying the possible tensors allowed by the symmetries. 
 \subsubsection{Dimension three}
For dimension three there are no covariant derivatives. The matrix elements of $\bar h s^\lambda h$ can only be proportional to $v^\lambda$. Since $v\cdot s=0$ we find that the matrix element is zero:
\begin{equation}
\dfrac1{2M_H}\langle H |\bar h s^\lambda h|H\rangle=0.
\end{equation}
\subsubsection{Dimension Four}
The matrix element of the operator $\bar h\, iD^{\mu_1}s^\lambda h$ can only be proportional to $\Pi^{\mu_1\lambda}$, since both $v^{\mu_1}=0$ and $v^\lambda=0$. But then for the choice $v=(1,0,0,0)$ the matrix element contains one space-like covariant derivative and is zero by parity. Thus  
\begin{equation}
\dfrac1{2M_H}\langle H |\bar h\, iD^{\mu_1}s^\lambda h|H\rangle=0.
\end{equation}
\subsubsection{Dimension Five}
The operator $\bar h\, iD^{\mu_1}iD^{\mu_2}s^\lambda h$ has three indices, all of which are orthogonal to $v$. As a result, we cannot use three $v$'s  or a product of one $\Pi$ and one $v$. There is only one possible structure:
\begin{equation}\label{SD5}
\dfrac1{2M_H}\langle H |\bar h\, iD^{\mu_1}iD^{\mu_2}s^\lambda h|H\rangle=i\tilde a^{(5)}\epsilon^{\rho\mu_1\mu_2\lambda}v_{\rho}. 
\end{equation}
The tensor $\epsilon^{\rho\mu_1\mu_2\lambda}v_{\rho}$ is antisymmetric under inversion as required.  
\subsubsection{Dimension Six}
There is only one possible tensor, a product of $v$ and $\epsilon$. Thus  
\begin{equation}\label{SD6}
\dfrac1{2M_H}\langle H |\bar h\, iD^{\mu_1}iD^{\mu_2}iD^{\mu_3}s^\lambda h|H\rangle=i\tilde a^{(6)}v^{\mu_2}\epsilon^{\rho\mu_1\mu_3\lambda}v_{\rho}. 
\end{equation}
Again the inversion symmetry is manifest.   
\subsubsection{Dimension Seven}
For the matrix elements of dimension seven spin-dependent operators there are  five independent tensors. One has 2 $v$'s and $\epsilon$ and four that have $\Pi$ and $\epsilon$. Thus 
\begin{eqnarray}\label{SD7}
&&\dfrac1{2M_H}\langle H |\bar h\, iD^{\mu_1}iD^{\mu_2}iD^{\mu_3}iD^{\mu_4}s^\lambda h|H\rangle=\nonumber\\
&&i\tilde a_{12}^{(7)}\left(\Pi^{\mu_1\mu_2}\epsilon^{\rho\mu_3\mu_4\lambda}v_{\rho}-\Pi^{\mu_4\mu_3}\epsilon^{\rho\mu_2\mu_1\lambda}v_{\rho}\right)+i\tilde a_{13}^{(7)}\left(\Pi^{\mu_1\mu_3}\epsilon^{\rho\mu_2\mu_4\lambda}v_{\rho}-\Pi^{\mu_4\mu_2}\epsilon^{\rho\mu_3\mu_1\lambda}v_{\rho}\right)+\nonumber\\
&+&i\tilde a_{14}^{(7)}\Pi^{\mu_1\mu_4}\epsilon^{\rho\mu_2\mu_3\lambda}v_{\rho}+i\tilde a_{23}^{(7)}\Pi^{\mu_2\mu_3}\epsilon^{\rho\mu_1\mu_4\lambda}v_{\rho}+i\tilde b^{(7)}v^{\mu_2}v^{\mu_3}\epsilon^{\rho\mu_1\mu_4\lambda}v_{\rho},
\end{eqnarray}
where we have imposed the inversion symmetry by combining  tensors in the second line of (\ref{SD7}) with the same non-perturbative parameters.

Naively it might seem that there are two other possible independent tensors that involve $\Pi^{\lambda\mu_i}$, namely $\Pi^{\mu_1\lambda}\epsilon^{\rho\mu_2\mu_3\mu_4}v_{\rho}-\Pi^{\mu_4\lambda}\epsilon^{\rho\mu_3\mu_2\mu_1}v_{\rho}$ and $\Pi^{\mu_2\lambda}\epsilon^{\rho\mu_1\mu_3\mu_4}v_{\rho}-\Pi^{\mu_3\lambda}\epsilon^{\rho\mu_4\mu_2\mu_1}v_{\rho}$. But this would be an over-counting. The tensor $\Pi^{\mu\nu}\epsilon^{\sigma\alpha\beta\rho}v_\sigma$ has five indices orthogonal to $v$, but in four space-time dimensions there can be only three different indices orthogonal to $v$. Since $\alpha\neq\beta\neq\rho$ and $\mu=\nu$, it follows that three of the indices in the set  $\left\{\alpha, \beta,\rho,\mu,\nu\right\}$ are equal. Therefore if $\lambda$ is equal to any $\mu_i$ it is also equal to some $\mu_j$ and hence $\mu_i=\mu_j$ and already included in the tensors of (\ref{SD7}). 

For the dimension seven spin-independent case one can construct operators with the same Lorentz structure but different color structure. We can check whether this is possible for the spin-dependent operators by contracting $\bar h\left\{[iD^{\mu_i},iD^{\mu_j}],[iD^{\mu_k},iD^{\mu_l}]\right\}h$ with the tensors on the right hand side of (\ref{SD7}). We find that all of these vanish, so there are no such operators. We will find the same result in section \ref{NRQCD_SD7}.

\subsubsection{Dimension Eight}\label{sec_SD8}
For the matrix elements of the dimension eight spin-dependent operators we can have one tensor with 3 $v$'s, $v^{\mu_2}v^{\mu_3}v^{\mu_4}\epsilon^{\rho\mu_1\mu_5\lambda}v_\rho$, and tensors which are of the form $v\,\Pi\, \epsilon$. Following the discussion above, the $\Pi$'s should depend only on $\mu_i$. Once we fix $v^{\mu_i}$ to be $v^{\mu_2},v^{\mu_3}$, or $v^{\mu_4}$, there are four indices left, which gives six pairs $\{\mu_j,\mu_k\}$ for $\Pi$. Including the constraints from inversion symmetry, we find   
\begin{eqnarray}\label{SD8}
&&\dfrac1{2M_H}\langle H |\bar h\, iD^{\mu_1}iD^{\mu_2}iD^{\mu_3}iD^{\mu_4}iD^{\mu_5}s^\lambda h|H\rangle=\nonumber\\
&&i\tilde a_{12}^{(8)}\left(v^{\mu_3}\Pi^{\mu_1\mu_2}\epsilon^{\rho\mu_4\mu_5\lambda}v_{\rho}-v^{\mu_3}\Pi^{\mu_4\mu_5}\epsilon^{\rho\mu_2\mu_1\lambda}v_{\rho}\right)+
i\tilde a_{14}^{(8)}\left(v^{\mu_3}\Pi^{\mu_1\mu_4}\epsilon^{\rho\mu_2\mu_5\lambda}v_{\rho}-v^{\mu_3}\Pi^{\mu_5\mu_2}\epsilon^{\rho\mu_4\mu_1\lambda}v_{\rho}\right)+\nonumber\\
&+&i\tilde a_{15}^{(8)}v^{\mu_3}\Pi^{\mu_1\mu_5}\epsilon^{\rho\mu_2\mu_4\lambda}v_{\rho}+i\tilde a_{24}^{(8)}v^{\mu_3}\Pi^{\mu_2\mu_4}\epsilon^{\rho\mu_1\mu_5\lambda}v_{\rho}+\nonumber\\
&+&i\tilde b_{13}^{(8)}\left(v^{\mu_2}\Pi^{\mu_1\mu_3}\epsilon^{\rho\mu_4\mu_5\lambda}v_{\rho}-v^{\mu_4}\Pi^{\mu_5\mu_3}\epsilon^{\rho\mu_2\mu_1\lambda}v_{\rho}\right)+
i\tilde b_{14}^{(8)}\left(v^{\mu_2}\Pi^{\mu_1\mu_4}\epsilon^{\rho\mu_3\mu_5\lambda}v_{\rho}-v^{\mu_4}\Pi^{\mu_5\mu_2}\epsilon^{\rho\mu_3\mu_1\lambda}v_{\rho}\right)+\nonumber\\
&+&i\tilde b_{15}^{(8)}\left(v^{\mu_2}\Pi^{\mu_1\mu_5}\epsilon^{\rho\mu_3\mu_4\lambda}v_{\rho}-v^{\mu_4}\Pi^{\mu_1\mu_5}\epsilon^{\rho\mu_3\mu_2\lambda}v_{\rho}\right)+
i\tilde b_{34}^{(8)}\left(v^{\mu_2}\Pi^{\mu_3\mu_4}\epsilon^{\rho\mu_1\mu_5\lambda}v_{\rho}-v^{\mu_4}\Pi^{\mu_3\mu_2}\epsilon^{\rho\mu_5\mu_1\lambda}v_{\rho}\right)+\nonumber\\
&+&i\tilde b_{35}^{(8)}\left(v^{\mu_2}\Pi^{\mu_3\mu_5}\epsilon^{\rho\mu_1\mu_4\lambda}v_{\rho}-v^{\mu_4}\Pi^{\mu_3\mu_1}\epsilon^{\rho\mu_5\mu_2\lambda}v_{\rho}\right)+
i\tilde b_{45}^{(8)}\left(v^{\mu_2}\Pi^{\mu_4\mu_5}\epsilon^{\rho\mu_1\mu_3\lambda}v_{\rho}-v^{\mu_4}\Pi^{\mu_2\mu_1}\epsilon^{\rho\mu_5\mu_3\lambda}v_{\rho}\right)+\nonumber\\
&+&i\tilde c^{(8)}v^{\mu_2}v^{\mu_3}v^{\mu_4}\epsilon^{\rho\mu_1\mu_5\lambda}v_{\rho}. 
\end{eqnarray}

As for the spin-independent case we can check if there are operators with the same Lorentz structure but different color structure by contracting  $\bar h\left\{[iD^{\mu_i},iD^{\mu_j}],[iD^{\mu_k},[iD^{\mu_l},iD^{\mu_m}]]\right\}h$, $\bar h\left\{[iD^{\mu_i},iD^{\mu_j}],[iD^{\mu_m},[iD^{\mu_k},iD^{\mu_l}]]\right\}h$, and $\bar h\left\{iD^{\mu_m},\left\{[iD^{\mu_i},iD^{\mu_j}],[iD^{\mu_k},iD^{\mu_l}]\right\}\right\}h$ with the tensors of the right hand side of (\ref{SD8}). We find six linearly-independent combinations, indicating that there will be six operators with two possible color structures. We will find the same result in section \ref{NRQCD_SD8}.  Including these possible color structures, there seventeen NRQCD (HQET) operators in total. 

\section{HQET Operators }\label{HQET}
We can now easily relate the matrix elements analyzed in the previous section to the HQET parameters listed in \cite{Mannel:2010wj}. As described in section \ref{General.1}, the operators listed \cite{Mannel:2010wj} relevant to tree level matching of power corrections to inclusive $B$-decays have color structure of the form $T^aT^b$ and not $\delta^{ab}$. The different color structure is generated via (at least) one gluon exchange.  We list the other operators in section  \ref{NRQED}.  
\subsection{Spin-independent operators}
\subsubsection{Dimension Five}
In \cite{Mannel:1994kv} the spin-independent matrix element is defined as 
\begin{equation}
\dfrac1{2M_H}\langle H(v)|\bar Q_v\, iD^{\mu_1}iD^{\mu_2}Q_v|H(v)\rangle=\dfrac13\lambda_1\,\Pi^{\mu_1\mu_2},
\end{equation}
while  \cite{Mannel:2010wj} defines
\begin{equation}
\dfrac1{2M_B}\langle B|\bar b_v\, iD^{\mu_1}iD^{\mu_2}b_v|B\rangle\,\Pi_{\mu_1\mu_2}=-\mu_\pi^2.
\end{equation}
Comparing to (\ref{SI5}) and using $\Pi^{\mu_1\mu_2}\Pi_{\mu_1\mu_2}=3$ we find that $-\mu_\pi^2=\lambda_1=3a^{(5)}$. One should keep in mind though that $\mu_\pi^2$ is not defined in the heavy quark limit, but using the full QCD $b$ fields \cite{Mannel:2010wj}. Therefore the relation between $-\mu_\pi^2$ and $\lambda_1$  has $1/m^2_b$ corrections.  
\subsubsection{Dimension Six}
In \cite{Mannel:1994kv} the spin-independent matrix element is defined as
\begin{equation}
\dfrac1{2M_H}\langle H(v)|\bar Q_v\, iD^{\mu_1}iD^{\mu_2}iD^{\mu_3}Q_v|H(v)\rangle=\dfrac13\rho_1\Pi^{\mu_1\mu_3}v^{\mu_2}.
\end{equation}
while  \cite{Mannel:2010wj} defines
\begin{equation}
\dfrac1{2M_B}\langle B|\bar h\, \Big[iD^{\mu_1},\big[iD^{\mu_2},iD^{\mu_3}\big]\Big]h|B\rangle\dfrac12\Pi_{\mu_1\mu_3}v_{\mu_2}=\rho_D^3.
\end{equation}
Comparing to (\ref{SI6}) we find that $\rho_D^3=\rho_1=3a^{(6)}$. 
\subsubsection{Dimension Seven}
In \cite{Mannel:2010wj} the four spin-independent matrix element are defined as
\begin{eqnarray}
2M_B\, m_1 &=& \langle B | \bar b_v \, i D_\rho i D_\sigma i D_\lambda iD_\delta 
\,b_v | B \rangle\,\, \scalebox{1.1}{$\frac{1}{3}$} 
\left(\Pi^{\rho \sigma}\Pi^{\lambda \delta} + 
\Pi^{\rho \lambda}\Pi^{\sigma \delta}+\Pi^{\rho \delta}\Pi^{\sigma \lambda}\right) \nonumber\\
\nonumber
2M_B\, m_2 &=& \langle B | \bar b_v \, \big[ i D_\rho ,  i D_\sigma \big] 
\big[ i D_\lambda ,  i D_\delta\big] \,b_v | B \rangle \:\Pi^{\rho\delta}v^\sigma 
v^\lambda \\
\nonumber
2M_B\, m_3 &=& \langle B | \bar b_v \, \big[i D_\rho ,  i D_\sigma\big]
\big[ iD_\lambda, iD_\delta\big] \,b_v|B\rangle\:\Pi^{\rho \lambda}\Pi^{\sigma \delta} \\
2M_B\, m_4 &=& \langle B | \bar b_v \, \Big\lbrace i D_\rho , \Big[i D_\sigma ,
\big[ i D_\lambda , i D_\delta\big]\Big]\Big \rbrace \,b_v | B \rangle \:
\Pi^{\sigma \lambda}\Pi^{\rho \delta}
\end{eqnarray}
Using the definitions of (\ref{SI7}) we find 
\begin{equation}
m_1=5\left[a_{12}^{(7)}+a_{13}^{(7)}+a_{14}^{(7)}\right],\,m_2=3b^{(7)},\, m_3=12\left[a_{13}^{(7)}-a_{14}^{(7)}\right],\,m_4=12\left[a_{12}^{(7)}-2a_{13}^{(7)}+a_{14}^{(7)}\right].
\end{equation}

\subsubsection{Dimension Eight}
In \cite{Mannel:2010wj} seven spin-independent matrix elements are listed\footnote{The change $b_v\to b$ is presumably a typo in \cite{Mannel:2010wj}.} as:
 \begin{eqnarray}
\nonumber
2M_B r_1 &=& \langle B | \bar b \,i  D_\rho\, (i v \cdot D)^3\, i  D^\rho \, b | B \rangle \\
\nonumber 
2M_B r_2 &=& \langle B | \bar b \,i  D_\rho\, (i v \cdot D)\, i  D^\rho\, i  D_\sigma\, i  D^\sigma \, b | B \rangle \\
\nonumber
2M_B r_3 &=& \langle B | \bar b \,i  D_\rho\, (i v \cdot D)\, i  D_\sigma\, i
D^\rho\, i  D^\sigma \, b | B \rangle \\  
\nonumber
2M_B r_4 &=& \langle B | \bar b \,i  D_\rho\, (i v \cdot D)\, i  D_\sigma\, i
D^\sigma\, i  D^\rho \, b | B \rangle \\  
\nonumber
2M_B r_5 &=& \langle B | \bar b \,i  D_\rho\, i  D^\rho\,(i v \cdot D)\,  i
D_\sigma\, i  D^\sigma \, b | B \rangle \\  
\nonumber
2M_B r_6 &=& \langle B | \bar b \,i  D_\rho\, i  D_\sigma\, (i v \cdot D)\, i
D^\sigma\, i  D^\rho \, b | B \rangle \\  
2M_B r_7 &=& \langle B | \bar b \,i  D_\rho\, i  D_\sigma\, (i v \cdot D)\, i
D^\rho\, i  D^\sigma \, b | B \rangle
\end{eqnarray}
Notice that for $r_2, r_3$, and $r_4$ the operators are not Hermitian.  In the following we assume that the contractions of indices is done using $\Pi$, i.e. the contracted indices are space-like. If they are contracted using the regular metric tensor, it slightly changes the relation of $r_4$ and $r_6$ to our basis. 
As written above, we can relate these matrix elements to the parameters introduced in (\ref{SI8}). We find\footnote{If we use the regular metric tensor for contractions, $r_4\to r_4+3c^{(8)}$ and $r_6\to r_6+3c^{(8)}$.}
\begin{eqnarray}
&&r_1=3c^{(8)},\nonumber\\
&&r_2=3\left[3a_{12}^{(8)}+a_{13}^{(8)}+a_{15}^{(8)}\right],\, r_3=3\left[a_{12}^{(8)}+3a_{13}^{(8)}+a_{15}^{(8)}\right],\, r_4=3\left[a_{12}^{(8)}+a_{13}^{(8)}+3a_{15}^{(8)}\right] \nonumber\\
&&r_5=3\left[3b_{12}^{(8)}+b_{14}^{(8)}+b_{15}^{(8)}\right],\, r_6=3\left[b_{12}^{(8)}+b_{14}^{(8)}+3b_{15}^{(8)}\right],\, r_7=3\left[b_{12}^{(8)}+3b_{14}^{(8)}+b_{15}^{(8)}\right].
\end{eqnarray}

\subsection{Spin-dependent operators}
Matrix elements of spin dependent operators of dimension three and four are zero. The first non-vanishing matrix element is of dimension five. 
\subsubsection{Dimension Five}
In \cite{Mannel:1994kv} the spin-dependent matrix element is defined as
\begin{equation}
\dfrac1{2M_H}\langle H(v)|\bar Q_v\, iD^{\mu_1}iD^{\mu_2}s^\lambda Q_v|H(v)\rangle=\dfrac12\lambda_2\,i\epsilon^{\rho\mu_1\mu_2\lambda}v_\rho,
\end{equation}
while  \cite{Mannel:2010wj} defines
\begin{equation}
\dfrac1{2M_B}\langle B|\bar b_v\, \dfrac12[iD^{\mu_1},iD^{\mu_2}](-i\sigma^{\mu_1\mu_2})b_v|B\rangle\,\Pi_{\mu_1\mu_2}=\mu_G^2.
\end{equation}
The matrix $(-i\sigma^{\mu\nu})$ is related to $s^\lambda$ via \cite{Mannel:1994kv} 
\begin{equation}
(-i\sigma^{\mu\nu})\to \frac{1+\vslash}{2}(-i\sigma^{\mu\nu})\frac{1+\vslash}{2}=iv_\alpha\epsilon^{\alpha\mu\nu\beta}s_\beta. 
\end{equation}
Comparing to (\ref{SD5})  we find that $\mu_G^2=3\lambda_2=6\tilde{a}^{(5)}$. As for $\mu_\pi^2$,  there are $1/m^2_b$ corrections to the relation between $\mu_G^2$ and $\lambda_2$.
\subsubsection{Dimension Six}
In \cite{Mannel:1994kv} the spin-dependent matrix element is defined as
\begin{equation}
\dfrac1{2M_H}\langle H(v)|\bar Q_v\, iD^{\mu_1}iD^{\mu_2}iD^{\mu_3}s^\lambda Q_v|H(v)\rangle=\dfrac12\rho_2\,iv_\nu\epsilon^{\nu\mu_1\mu_3\lambda}v^{\mu_2},
\end{equation}
while  \cite{Mannel:2010wj} defines
\begin{equation}
\dfrac1{2M_B}\langle B|\bar b_v\,\frac12\big\{iD^{\mu_1},\left [iD^{\mu_2},iD^{\mu_3}\right]\big\}(-i\sigma^{\alpha\beta})b_v|B\rangle\,\Pi_{\mu_1\alpha}\Pi_{\mu_3\beta}v_{\mu_2}=\rho_{LS}^3.
\end{equation}
Comparing to (\ref{SD6})  we find that $\rho_{LS}^3=3\rho_2=6\tilde{a}^{(6)}$. 
\subsubsection{Dimension Seven}
In \cite{Mannel:2010wj} the five spin-dependent matrix element are defined as
\begin{eqnarray}
2M_B\, m_5 &=& \langle B | \bar b_v \, \big[ i D_\rho  , i D_\sigma\big] 
\big[ i D_\lambda ,  i D_\delta\big]  \big (-i \sigma_{\alpha \beta}\big)\,b_v
| B \rangle \,\,  \Pi^{\alpha \rho} \Pi^{\beta \delta} v^\sigma v^\lambda \nonumber\\
\nonumber
2M_B\, m_6 &=& \langle B | \bar b_v \, \big[ i D_\rho  , i D_\sigma \big] 
\big[i D_\lambda , i D_\delta\big]  \big (-i \sigma_{\alpha \beta}\big)\,b_v 
| B \rangle \,\, \Pi^{\alpha \sigma} \Pi^{\beta \lambda} \Pi^{\rho \delta}  \nonumber\\
2M_B\, m_7 &=& \langle B | \bar b_v \,\Big \lbrace \big \lbrace i D_\rho,  
i D_\sigma \big \rbrace ,\big[ i D_\lambda , i D_\delta \big] \Big\rbrace 
\big (-i \sigma_{\alpha \beta}\big)\,b_v | B \rangle \,\, \Pi^{\sigma \lambda}
\Pi^{\alpha \rho} \Pi^{\beta \delta} \nonumber\\
2M_B\, m_8 &=&  \langle B | \bar b_v \,\Big \lbrace \big \lbrace
i D_\rho ,  i D_\sigma \big \rbrace, \big[i D_\lambda  , i D_\delta \big] 
\Big\rbrace \big (-i \sigma_{\alpha \beta}\big)\,b_v | B \rangle \: 
\Pi^{\rho \sigma } \Pi^{\alpha \lambda} \Pi^{\beta \delta} \nonumber \\
2M_B\, m_9 &=& \langle B | \bar b_v \, \bigg[ i D_\rho ,  \Big[ i D_\sigma ,
\big[ i D_\lambda ,  i D_\delta \big]\Big]\bigg]  \big (-i \sigma_{\alpha \beta}\big) 
\,b_v | B \rangle \,\, \Pi^{\rho \beta} \Pi^{\lambda \alpha} \Pi^{\sigma \delta}\,.
\end{eqnarray}
Comparing to (\ref{SD7})  we find
\begin{eqnarray}
&&m_5=6\tilde{b}^{(7)},\,m_6=6\left[-2 \tilde{a}^{(7)}_{13} + \tilde{a}^{(7)}_{14} +\tilde{a}^{(7)}_{23}\right],\,m_7=12\left[4 \tilde{a}^{(7)}_{12} -3 \tilde{a}^{(7)}_{14} +3\tilde{a}^{(7)}_{23}\right]\nonumber\\
&&m_8=48\left[3 \tilde{a}^{(7)}_{12} - \tilde{a}^{(7)}_{14} +\tilde{a}^{(7)}_{23}\right],\,m_9=12\left[5 \tilde{a}^{(7)}_{12}-4 \tilde{a}^{(7)}_{14} -3 \tilde{a}^{(7)}_{14} +2\tilde{a}^{(7)}_{23}\right].
\end{eqnarray}

\subsubsection{Dimension Eight}
In \cite{Mannel:2010wj} eleven spin-dependent matrix elements are listed as:
 \begin{eqnarray}
\nonumber
2M_B r_{8} &=& \langle B | \bar b \,i   D_\mu \, (i v \cdot D)^3\, i   D_\nu
\, (-i \sigma^{\mu \nu })\,b | B \rangle \\  
\nonumber
2M_B r_{9} &=& \langle B | \bar b \,i   D_\mu \, (i v \cdot D)\, i   D_\nu
\, i   D_\rho\, i   D^\rho \,(-i \sigma^{\mu \nu })\, b | B \rangle \\  
\nonumber
2M_B r_{10} &=& \langle B | \bar b \,i   D_\rho\, (i v \cdot D)\, i
D^\rho\, i   D_\mu \, i   D_\nu  \,(-i \sigma^{\mu \nu })\, b | B \rangle \\
\nonumber
2M_B r_{11} &=& \langle B | \bar b \,i   D_\rho\, (i v \cdot D)\, i   D_\mu
\, i   D^\rho\, i   D_\nu  \,(-i \sigma^{\mu \nu })\, b | B \rangle \\  
\nonumber
2M_B r_{12} &=& \langle B | \bar b \,i   D_\mu \, (i v \cdot D)\, i
D_\rho\, i   D_\nu \, i   D^\rho \,(-i \sigma^{\mu \nu })\, b | B \rangle \\
\nonumber
2M_B r_{13} &=& \langle B | \bar b \,i   D_\rho\, (i v \cdot D)\, i   D_\mu
\, i   D_\nu \, i   D^\rho \,(-i \sigma^{\mu \nu })\, b | B \rangle \\  
\nonumber
2M_B r_{14} &=& \langle B | \bar b \,i   D_\mu \, (i v \cdot D)\, i
D_\rho\, i   D^\rho\, i   D_\nu  \,(-i \sigma^{\mu \nu })\, b | B \rangle \\
\nonumber
2M_B r_{15} &=& \langle B | \bar b \,i   D_\mu \, i   D_\nu \, (i v \cdot
D)\, i   D_\rho\, i   D^\rho \,(-i \sigma^{\mu \nu })\, b | B \rangle \\  
\nonumber
2M_B r_{16} &=& \langle B | \bar b \,i   D_\rho\, i   D_\mu \, (i v \cdot
D)\, i   D_\nu \, i   D^\rho \,(-i \sigma^{\mu \nu })\, b | B \rangle \\  
\nonumber
2M_B r_{17} &=& \langle B | \bar b \,i   D_\mu \, i   D_\rho\, (i v \cdot
D)\, i   D^\rho\, i   D_\nu  \,(-i \sigma^{\mu \nu })\, b | B \rangle \\  
2M_B r_{18} &=& \langle B | \bar b \,i   D_\rho\, i   D_\mu \, (i v \cdot D)\, i   D^\rho\, i   D_\nu  \,(-i \sigma^{\mu \nu })\, b | B \rangle \,.
\end{eqnarray}
Notice that for $r_9-r_{14}$ the operators are not Hermitian.  In the following we assume that the contractions of indices is done using $\Pi$, i.e. the contracted indices are space-like. We can relate these matrix elements to the parameters introduced in (\ref{SD8}). We find
\begin{eqnarray}
&&r_8=6\tilde{c}^{(8)}\nonumber\\
&&r_9=-6\left[\tilde{b}_{14}^{(8)}+\tilde{b}_{15}^{(8)}-\tilde{b}_{34}^{(8)}-\tilde{b}_{35}^{(8)}-3\tilde{b}_{45}^{(8)}\right],\,
r_{10}=6\left[3\tilde{b}_{13}^{(8)}+\tilde{b}_{14}^{(8)}-\tilde{b}_{15}^{(8)}+\tilde{b}_{34}^{(8)}-\tilde{b}_{35}^{(8)}\right],\,\nonumber\\
&&r_{11}=6\left[\tilde{b}_{13}^{(8)}+3\tilde{b}_{14}^{(8)}+\tilde{b}_{15}^{(8)}+\tilde{b}_{34}^{(8)}-\tilde{b}_{45}^{(8)}\right],\, 
r_{12}=6\left[-\tilde{b}_{13}^{(8)}+\tilde{b}_{15}^{(8)}+\tilde{b}_{34}^{(8)}+3\tilde{b}_{35}^{(8)}+\tilde{b}_{45}^{(8)}\right],\, \nonumber\\
&&r_{13}=-6\left[\tilde{b}_{13}^{(8)}-\tilde{b}_{14}^{(8)}-3\tilde{b}_{15}^{(8)}-\tilde{b}_{35}^{(8)}+\tilde{b}_{45}^{(8)}\right],\, 
r_{14}=6\left[\tilde{b}_{13}^{(8)}+\tilde{b}_{14}^{(8)}+3\tilde{b}_{34}^{(8)}+\tilde{b}_{35}^{(8)}+\tilde{b}_{45}^{(8)}\right],\, \nonumber\\
&&r_{15}=6\left[3\tilde{a}_{12}^{(8)}-\tilde{a}_{15}^{(8)}+3\tilde{a}_{24}^{(8)}\right],\, 
r_{16}=6\left[-2\tilde{a}_{12}^{(8)}+2\tilde{a}_{14}^{(8)}+3\tilde{a}_{15}^{(8)}\right],\, \nonumber\\
&&r_{17}=6\left[2\tilde{a}_{12}^{(8)}+2\tilde{a}_{14}^{(8)}+3\tilde{a}_{24}^{(8)}\right],\, 
r_{18}=6\left[3\tilde{a}_{14}^{(8)}+\tilde{a}_{15}^{(8)}+\tilde{a}_{24}^{(8)}\right].
\end{eqnarray}

\section{NRQED and NRQCD operators}\label{NRQED}
We now relate the known NRQED and NRQCD operators up to dimension eight to the decomposition of section \ref{General}. As we will see, there are dimension eight  NRQCD operators that do not appear in the dimension eight NRQED Lagrangian. These operators were not considered before in the literature. We will list them  in section \ref{Applications}. 

The $1/M^3$  NRQCD Lagrangian that contains operators up to dimension seven was given in \cite{Manohar:1997qy}
\begin{eqnarray}\label{LNRQCD3}
&&{\cal L}_{\mbox{\scriptsize NRQCD}}^{\mbox{\scriptsize dim$\leq$7}} = \psi^\dagger
  \bigg\{  i D_t  + c_2{\bm{D}^2 \over 2 M}+ 
  c_F g{ \bm{\sigma}\cdot \bm{B} \over 2M}   
+ c_D g{\bm{D\cdot E}-\bm{E\cdot D} \over 8 M^2}  + i c_S g{ \bm{\sigma}
    \cdot ( \bm{D} \times \bm{E} - \bm{E}\times \bm{D} ) \over 8 M^2}+  \nonumber\\
    &&+ c_4{\bm{D}^4 \over 8 M^3} + i c_M g { \{ \bm{D}^i , ( \bm{D} \times \bm{B} - \bm{B}\times \bm{D} )^i \}\over 8 M^3}   + c_{W1}g {  \{ \bm{D}^2 ,  \bm{\sigma}\cdot \bm{B} \}  \over 8 M^3} - c_{W2}g {  \bm{D}^i \bm{\sigma}\cdot
    \bm{B} \bm{D}^i \over 4 M^3 }+ \nonumber\\
    && + c_{p^\prime p} g { \bm{\sigma} \cdot
    \bm{D} \bm{B}\cdot \bm{D} + \bm{D}\cdot\bm{B} \bm{\sigma}\cdot \bm{D}
    \over  8 M^3} +   c_{A1} g^2\,{ \left(\bm{B}_a^i\bm{B}_b^i - \bm{E}_a^i\bm{E}_b^i\right)\,T^aT^b \over 8 M^3} - c_{A2} g^2\,{\bm{E}_a^i\bm{E}_b^i\,T^aT^b
    \over 16 M^3 }+    \nonumber\\ 
    &&+  c_{A3} g^2\,{ \left(\bm{B}_a^i\bm{B}_b^i - \bm{E}_a^i\bm{E}_b^i\right)\delta^{ab} \over 8 M^3} - c_{A4} g^2\,{\bm{E}_a^i\bm{E}_b^i\,\delta^{ab}
    \over 16 M^3 }\nonumber\\
  &&
     -c_{B1} g^2 {\bm{\sigma\cdot} (\bm{B}_a\bm{\times B}_b - \bm{E}_a\times\bm{E}_b)f^{abc}T^c \over 16 M^3}+ c_{B2} g^2 {\bm{\sigma\cdot} (\bm{E}_a\times\bm{E}_b)f^{abc}T^c \over 16 M^3}
       \bigg\} \psi.
   \end{eqnarray}
We follow the notation of  \cite{Manohar:1997qy}, but display explicitly the color factors for terms bilinear in $\bm{E}$ or $\bm{B}$.  For terms linear in $\bm{E}$ or $\bm{B}$ we have $\bm{E}^i\equiv\bm{E}^i_aT^a$ and $\bm{B}^i\equiv\bm{B}^i_aT^a$. 
The operators on the last line of (\ref{LNRQCD3}) appear only for NRQCD and not NRQED. Also, for NRQED the operators whose coefficients are $c_{A1}$ and $c_{A3}$ ($c_{A2}$ and $c_{A4}$) are identical.

The $1/M^4$  NRQED Lagrangian that contains operators of dimension eight was given in \cite{Hill:2012rh}
\begin{align}\label{LNRQED4}
&{\cal L}_{\mbox{\scriptsize NRQED}}^{\mbox{\scriptsize dim$=$8}} = \psi^\dagger
  \bigg\{ 
c_{X1}g { [ \bm{D}^2 , \bm{D}\cdot \bm{E} + \bm{E}\cdot\bm{D} ] \over M^4 }
+ c_{X2}g { \{ \bm{D}^2 , [\bm{\partial}\cdot\bm{E}] \} \over M^4 }
+ c_{X3}g { [\bm{\partial}^2 \bm{\partial}\cdot\bm{E}] \over M^4 } \nonumber\\
&\quad 
+ i c_{X4}g^2 { \{ \bm{D}^i , [\bm{E}\times\bm{B}]^i \} \over M^4 } 
+ ic_{X5} g { \bm{D}^i \bm{\sigma}\cdot ( \bm{D}\times\bm{E} - \bm{E}\times\bm{D} )\bm{D}^i   \over M^4} 
+ ic_{X6} g { \epsilon^{ijk} \sigma^i \bm{D}^j [\bm{\partial}\cdot\bm{E}] \bm{D}^k \over M^4} 
\nonumber\\
&\quad
+ c_{X7} g^2 { \bm{\sigma}\cdot\bm{B} [\bm{\partial}\cdot\bm{E}] \over M^4} 
+ c_{X8} g^2 { [\bm{E}\cdot\bm{\partial} \bm{\sigma}\cdot\bm{B} ] \over M^4}
+ c_{X9} g^2 { [\bm{B}\cdot\bm{\partial} \bm{\sigma}\cdot\bm{E} ] \over M^4} 
\nonumber\\
&\quad
+ c_{X10} g^2 { [\bm{E}^i \bm{\sigma}\cdot\bm{\partial} \bm{B}^i] \over M^4}
+ c_{X11} g^2 { [\bm{B}^i \bm{\sigma}\cdot\bm{\partial} \bm{E}^i] \over M^4} 
+ c_{X12} g^2 { \bm{\sigma}\cdot \bm{E}\times [{\partial_t}\bm{E}-\bm{\partial}\times\bm{B} ] \over M^4} \bigg\} \psi  \,.
\end{align}
Some of these operators need to be rewritten in a form appropriate for NRQCD operators, e.g.  not assuming that $\bm{E}$ and $\bm{B}$ commute.  We will do that below. 

The general procedure we will follow is to take a  general NRQCD (NRQED) operator of the form $\psi^\dagger O\psi$ where $O$ is written in terms of $\bm {D},\bm{E},\bm{B}$. We change $\psi\to h$ and $\psi^\dagger\to \bar h$ and write $O$ in terms of covariant derivatives $iD^\mu$ contracted with $\Pi$ and $v$. The matrix element of the resulting operator can be written in terms of the  parameters of section \ref{General}. The utility of this method is that given two NRQCD operators we can immediately determine if they are linearly independent, based on the linear combination of parameters that corresponds to each operator. Possible multiple color factors for operators with the same Lorentz structure are considered separately.  We will illustrate this procedure in detail below. 

\subsection{Spin-independent operators}
\subsubsection{Dimension Four}
There is one spin-independent operator of dimension four in (\ref{LNRQCD3}). It has one time-like covariant  derivative $\psi^\dagger iD_t\psi$. The corresponding HQET operator is $\bar h iv\cdot Dh$ whose matrix element vanishes. 
\subsubsection{Dimension Five}
There is one spin-independent operator of dimension five in (\ref{LNRQCD3}): $\psi^\dagger\bm{D^2}\psi$ . The operator $\bm{D^2}$ can be written as $\Pi_{\mu_1\mu_2}iD^{\mu_1}iD^{\mu_2}$. Changing $\psi\to h$ and $\psi^\dagger\to \bar h$, we get 
\begin{equation}
\psi^\dagger\bm{D^2}\psi\to \dfrac1{2M_H}\langle H |\bar h\, iD^{\mu_1}iD^{\mu_2}\Pi_{\mu_1\mu_2}h|H\rangle=3a^{(5)}.
\end{equation}
\subsubsection{Dimension Six}
There is one spin-independent operator of dimension six  in (\ref{LNRQCD3}): $g\psi^\dagger \left(\bm{D\cdot E}-\bm{E\cdot D}\right)\psi$. It can be written as $-v_{\mu_2}\Pi_{\mu_1\mu_3}\left[iD^{\mu_1},\left[iD^{\mu_2},iD^{\mu_3}\right]\right]$. Changing $\psi\to h$ and $\psi^\dagger\to \bar h$, we get 
\begin{equation}
-\psi^\dagger v_{\mu_2}\Pi_{\mu_1\mu_3}\left[iD^{\mu_1},\left[iD^{\mu_2},iD^{\mu_3}\right]\right]\psi\to -\dfrac1{2M_H}\langle H |\bar h v_{\mu_2}\Pi_{\mu_1\mu_3}\left[iD^{\mu_1},\left[iD^{\mu_2},iD^{\mu_3}\right]\right]h|H\rangle=-6a^{(6)}.
\end{equation}
\subsubsection{Dimension Seven}\label{NRQCD_SI7}
There are six spin-independent dimension-seven operators in (\ref{LNRQCD3}): $\psi^\dagger\bm{D^4}\psi$, $g\psi^\dagger\{ \bm{D}^i , ( \bm{D} \times \bm{B} - \bm{B}\times \bm{D} )^i \}\psi$, $g^2\psi^\dagger\left(\bm{B}_a^i\bm{B}_b^i - \bm{E}_a^i\bm{E}_b^i\right)\,T^aT^b \psi$, $-g^2\psi^\dagger\bm{E}_a^i\bm{E}_b^i\,T^aT^b\psi$, $g^2\psi^\dagger\left(\bm{B}_a^i\bm{B}_b^i - \bm{E}_a^i\bm{E}_b^i\right)\,\delta^{ab} \psi$, $-g^2\psi^\dagger\bm{E}_a^i\bm{E}_b^i\,\delta^{ab}\psi$. Changing $\psi\to h$ and $\psi^\dagger\to \bar h$ the matrix elements of these operators are 
\begin{eqnarray}
&&\psi^\dagger \bm{D^4} \psi\to\dfrac1{2M_H}\langle H |\bar h\,  iD^{\mu_1}iD^{\mu_2} iD^{\mu_3}iD^{\mu_4} h|H\rangle\Pi_{\mu_1\mu_2}\Pi_{\mu_3\mu_4}=3\left(3a^{(7)}_{12} + a^{(7)}_{13} + a^{(7)}_{14}\right),\nonumber\\
&&\psi^\dagger\, g\{ \bm{D}^i , ( \bm{D} \times \bm{B} - \bm{B}\times \bm{D} )^i \} \psi\to\dfrac1{2M_H}\langle H |\bar h\, \big\{ iD^{\mu_1},\left[iD^{\mu_2},\left[ iD^{\mu_3},iD^{\mu_4}\right]\,\right]\big\} h|H\rangle\Pi_{\mu_1\mu_4}\Pi_{\mu_2\mu_3}\nonumber\\&&=12\left(a^{(7)}_{12} -2 a^{(7)}_{13} + a^{(7)}_{14}\right),\nonumber\\
&&g^2\psi^\dagger\left(\bm{B}_a^i\bm{B}_b^i - \bm{E}_a^i\bm{E}_b^i\right)\,T^aT^b \psi,g^2\psi^\dagger\left(\bm{B}_a^i\bm{B}_b^i - \bm{E}_a^i\bm{E}_b^i\right)\,\delta^{ab} \psi\to\nonumber\\
&&-\frac12\dfrac1{2M_H}\langle H |\bar h\,\left[iD^{\mu_1},iD^{\mu_2}\right]\left[ iD^{\mu_3},iD^{\mu_4}\right] h|H\rangle g_{\mu_1\mu_3}g_{\mu_2\mu_4}=
3\left(-2a^{(7)}_{13} +2 a^{(7)}_{14} +b^{(7)}\right),
\nonumber
\end{eqnarray}
\begin{eqnarray}
&&-g^2\psi^\dagger\bm{E}_a^i\bm{E}_b^i\,T^aT^b\psi,\, -g^2\psi^\dagger\bm{E}_a^i\bm{E}_b^i\,\delta^{ab}\psi
\to-\dfrac1{2M_H}\langle H |\bar h\,\left[iD^{\mu_1},iD^{\mu_2}\right]\left[ iD^{\mu_3},iD^{\mu_4}\right] h|H\rangle g_{\mu_1\mu_3}v_{\mu_2}v_{\mu_4}\nonumber\\
&&=-3b^{(7)}.
\end{eqnarray}
These linear combinations of $a^{(7)}_{12},a^{(7)}_{13},a^{(7)}_{14}$ and $b^{(7)}$ are all independent of each other.  We also have two pairs of operators that differ only by their color structure. These indeed depend on the linear combinations we have identified earlier, namely $a_{13}^{(7)} - a_{14}^{(7)}$, and $b^{(7)}$.  
\subsubsection{Dimension Eight}\label{NRQCD_SI8}
There are  four spin-independent dimension-eight operators in the dimension-eight NRQED Lagrangian (\ref{LNRQED4}).  We rewrite three of them in a form appropriate for NRQCD. Thus we rewrite $g\psi^\dagger\{ \bm{D}^2 , [\bm{\partial}\cdot\bm{E}] \}\psi$ as $g\psi^\dagger \{ \bm{D}^2 , [\bm{D}^i, \bm{E}^i]\}\psi$, $g\psi^\dagger[\bm{\partial}^2 \bm{\partial}\cdot\bm{E}]\psi$ as $g\psi^\dagger[\bm{D}^i, [\bm{D}^i, [\bm{D}^j,\bm{E}^j]] ]\psi$. The operator  $g^2\psi^\dagger\{i \bm{D}^i , [\bm{E}\times\bm{B}]^i \} \psi$ can be generalized for NRQCD as either $\frac12g^2\psi^\dagger\{i \bm{D}^i , \epsilon^{ijk}\bm{E}_a^j\bm{B}_b^k\,\{T^a,T^b\}\}\psi$ or $g^2\psi^\dagger\{i \bm{D}^i , \epsilon^{ijk}\bm{E}_a^j\bm{B}_b^k\,\delta^{ab}\}\psi$. Replacing $\psi\to h$ and $\psi^\dagger\to \bar h$ the matrix elements of these operators are 
\begin{eqnarray}\label{NRQED8SI}
&&g\psi^\dagger  [ \bm{D}^2 ,  \{\bm{D}^i, \bm{E}^i\} ] \psi\to-\dfrac1{2M_H}\langle H |\bar h\,  [iD^{\mu_1}iD^{\mu_2},\{ iD^{\mu_3},[iD^{\mu_4},iD^{\mu_5}] \}]h|H\rangle v_{\mu_4}\Pi_{\mu_1\mu_2}\Pi_{\mu_3\mu_5}\nonumber\\
&&=-6\left(3b^{(8)}_{12} + b^{(8)}_{14} + b^{(8)}_{15}\right),\nonumber\\
&&g\psi^\dagger  \{ \bm{D}^2 , [\bm{D}^i, \bm{E}^i]\} \psi\to-\dfrac1{2M_H}\langle H |\bar h\,  \{iD^{\mu_1}iD^{\mu_2},[ iD^{\mu_3},[iD^{\mu_4},iD^{\mu_5}] ]\}h|H\rangle v_{\mu_4}\Pi_{\mu_1\mu_2}\Pi_{\mu_3\mu_5}\nonumber\\
&&=-6\left(6a^{(8)}_{12} +2 a^{(8)}_{13} + 2a^{(8)}_{15}-3b^{(8)}_{12} - b^{(8)}_{14} - b^{(8)}_{15}\right),\nonumber\\
&&\psi^\dagger\, g [\bm{D}^i, [\bm{D}^i, [\bm{D}^j,\bm{E}^j]]] \psi\to\nonumber\\
&&\to -\dfrac1{2M_H}\langle H |\bar h\,  [iD^{\mu_1},[iD^{\mu_2},[ iD^{\mu_3},[iD^{\mu_4},iD^{\mu_5}] ]]]h|B\rangle v_{\mu_4}\Pi_{\mu_1\mu_2}\Pi_{\mu_3\mu_5}\nonumber\\
&&=-6\left(8a^{(8)}_{12} +4 a^{(8)}_{13} + 8a^{(8)}_{15}-5b^{(8)}_{12} - 3b^{(8)}_{14} - 7b^{(8)}_{15}\right),\nonumber\\
&&\dfrac{g^2}{2}\psi^\dagger\,  \{i \bm{D}^i , \epsilon^{ijk}\bm{E}_a^j\bm{B}_b^k\,\{T^a,T^b\}\}\psi,\,g^2\psi^\dagger\,  \{i \bm{D}^i , \epsilon^{ijk}\bm{E}_a^j\bm{B}_b^k\,\delta^{ab}\}\psi\to\nonumber\\
&&\to\dfrac12\dfrac1{2M_H}\langle H |\bar h\,  \{iD^{\mu_1},\{[iD^{\mu_2}, iD^{\mu_3}],[iD^{\mu_4},iD^{\mu_5}] \}\}h|B\rangle v_{\mu_2}\Pi_{\mu_1\mu_4}\Pi_{\mu_3\mu_5}\nonumber\\
&&=6\left(a^{(8)}_{12} -a^{(8)}_{15}- b^{(8)}_{14} + b^{(8)}_{15}\right).
\end{eqnarray}
Notice that the last line of depends on the linear combination that was anticipated in section \ref{sec_SI8}. We anticipated eight dimension eight spin-independent operators  in section \ref{sec_SI8}, but we have listed only five so far. It is clear the NRQCD Lagrangian will contain three extra spin-independent operators.  We will list them in section \ref{NRQCD_New}. 
\subsection{Spin-dependent operators}
Our convention for sign of the Levi-Civita tensor is $\epsilon^{0123}=-1$ and $\epsilon_{0123}=1$. As a result, the three dimensional contraction $\epsilon_{ijk}A^iB^jC^k\to-\epsilon_{0\mu\nu\alpha}A^\mu B^\nu C^\mu$ in four dimensions, assuming $A^i\to A^\mu$ etc. The overall minus sign arises from the three space-like contractions. Since $D^\mu=\left(D^0,-\bm{D}\right)$, we have an extra minus sign for each space-like derivative that appears in the triple product.
\subsubsection{Dimension Five }
There are no dimension four spin-dependent operators. There is one spin-dependent operator of dimension five in (\ref{LNRQCD3}): $g\psi^\dagger\bm{\sigma\cdot B}\psi$. It can be written as $-\dfrac{i}{2}\psi^\dagger\epsilon^{ijk}\bm{\sigma}^i[i\bm{D}^j,i\bm{D}^k]\psi$. As explained above, $\epsilon^{ijk}\bm{\sigma}^i[i\bm{D}^j,i\bm{D}^k]\to -\epsilon_{\rho\lambda\mu_1\mu_2} v^\rho s^\lambda\, [iD^{\mu_1},iD^{\mu_2}]$. Changing $\psi\to h$, $\psi^\dagger\to \bar h$ we get 
\begin{equation}
g\psi^\dagger\bm{\sigma\cdot B}\psi\to \dfrac1{2M_H}\dfrac{1}{2}i \epsilon_{\rho\mu_1\mu_2\lambda}v^\rho \langle H |\bar h\,s^\lambda\,[iD^{\mu_1},iD^{\mu_2}]h|H\rangle=6\tilde{a}^{(5)}.
\end{equation}
\subsubsection{Dimension Six}
There is one dimension six spin-dependent operator  in (\ref{LNRQCD3}): $ig\,\psi^\dagger \bm{\sigma}
    \cdot ( \bm{D} \times \bm{E} - \bm{E}\times \bm{D})\psi $. The corresponding HQET matrix element is 
\begin{eqnarray}
&&ig\,\psi^\dagger \bm{\sigma}
    \cdot ( \bm{D} \times \bm{E} - \bm{E}\times \bm{D})\psi \to -\dfrac1{2M_H} i\epsilon_{\rho\lambda\mu_1\mu_3}v^\rho  v_{\mu_2}\,\langle H |\bar h\,s^\lambda\{iD^{\mu_1},[iD^{\mu_2},iD^{\mu_3}]\}h|H\rangle=\nonumber\\
&& =-12\tilde{a}^{(6)}.
\end{eqnarray}
\subsubsection{Dimension Seven}\label{NRQCD_SD7}
There are five spin-dependent operator in (\ref{LNRQCD3}):  $g \psi^\dagger  \{ \bm{D}^2 ,  \bm{\sigma}\cdot \bm{B} \}\psi$, $g\psi^\dagger \bm{D}^i \bm{\sigma}\cdot \bm{B} \bm{D}^i\psi$,  $g\psi^\dagger  \bm{\sigma} \cdot\bm{D} \bm{B}\cdot \bm{D} + \bm{D}\cdot\bm{B} \bm{\sigma}\cdot \bm{D}\psi$, $g^2\psi^\dagger \bm{\sigma\cdot} (\bm{B}_a\times\bm{B}_b)f^{abc}T^c \psi$, and $g^2\psi^\dagger \bm{\sigma\cdot}( \bm{E}_a\times\bm{E}_b)f^{abc}T^c \psi$.  
The last two appear in a different linear combination in (\ref{LNRQCD3}). Notice that there are no operators with the same Lorentz structure and multiple color structures as we anticipated before.  Changing $\psi\to h$, $\psi^\dagger\to \bar h$ we get 
\begin{eqnarray}
&&g \psi^\dagger \{ \bm{D}^2 ,  \bm{\sigma}\cdot \bm{B} \}\psi\to  \dfrac1{2M_H}\dfrac12i\epsilon_{\rho\mu_3\mu_4\lambda}v^\rho \Pi_{\mu_1\mu_2}\langle H |\bar h\,s^\lambda \{iD^{\mu_1}iD^{\mu_2},[iD^{\mu_3},iD^{\mu_4}]\}h|H\rangle=\nonumber\\
&&=12 \left(3 \tilde{a}_{12}^{(7)} -\tilde{a}_{14}^{(7)} + \tilde{a}_{23}^{(7)}\right),\nonumber\\
&&g\psi^\dagger \bm{D}^i \bm{\sigma}\cdot \bm{B} \bm{D}^i\psi\to  \dfrac1{2M_H}\dfrac12i\epsilon_{\rho\mu_2\mu_3\lambda}v^\rho \Pi_{\mu_1\mu_4}\langle H |\bar h\,s^\lambda iD^{\mu_1}[iD^{\mu_2},iD^{\mu_3}]iD^{\mu_4}\,h|H\rangle=\nonumber\\
&&=6 \left(-2 \tilde{a}_{12}^{(7)} +2\tilde{a}_{13}^{(7)} +3 \tilde{a}_{14}^{(7)}\right),\nonumber
\end{eqnarray}
\begin{eqnarray}
&&g\psi^\dagger  \bm{\sigma} \cdot\bm{D} \bm{B}\cdot \bm{D} + \bm{D}\cdot\bm{B} \bm{\sigma}\cdot \bm{D}\psi\to  -\dfrac1{2M_H}\dfrac12i\epsilon_{\rho\mu_1\mu_2\mu_3}v^\rho \Pi_{\lambda\mu_4}\langle H |\bar h\,s^\lambda iD^{\mu_1}[iD^{\mu_2},iD^{\mu_3}]iD^{\mu_4}\,h|H\rangle\nonumber\\
&&-\dfrac1{2M_H}\dfrac12i\epsilon_{\rho\mu_4\mu_2\mu_3}v^\rho \Pi_{\lambda\mu_1}\langle H |\bar h\,s^\lambda iD^{\mu_1}[iD^{\mu_2},iD^{\mu_3}]iD^{\mu_4}\,h|H\rangle=-12 \left( \tilde{a}_{12}^{(7)} -\tilde{a}_{13}^{(7)} + \tilde{a}_{14}^{(7)}\right),\nonumber\\
&&g^2\psi^\dagger \bm{\sigma\cdot} (\bm{B}_a\times\bm{B}_b)f^{abc}T^c \psi\to  \dfrac1{2M_H}\dfrac12i\epsilon_{\rho\mu_1\mu_2\mu_4}v^\rho \Pi_{\lambda\mu_3}\langle H |\bar h\,s^\lambda[iD^{\mu_1},iD^{\mu_2}][iD^{\mu_3},iD^{\mu_4}]\,h|H\rangle=\nonumber\\
&&=6 \left(2 \tilde{a}_{13}^{(7)} -\tilde{a}_{14}^{(7)} - \tilde{a}_{23}^{(7)}\right),\nonumber\\
&&g^2\psi^\dagger \bm{\sigma\cdot}( \bm{E}_a\times\bm{E}_b)f^{abc}T^c \psi\to  \dfrac1{2M_H}\i\epsilon_{\rho\mu_2\mu_4\lambda}v^\rho v_{\mu_1}v_{\mu_3}\langle H |\bar h\,s^\lambda[iD^{\mu_1},iD^{\mu_2}][iD^{\mu_3},iD^{\mu_4}]\,h|H\rangle=\nonumber\\
&&=-6 \tilde{b}^{(7)}.
\end{eqnarray}
\subsubsection{Dimension Eight}\label{NRQCD_SD8}
There are eight spin-dependent dimension-eight operators in the $1/M^4$ NRQED Lagrangian (\ref{LNRQED4}).  For the NRQCD operators we rewrite $\psi^\dagger\epsilon^{ijk} \sigma^i \bm{D}^j [\bm{\partial}\cdot\bm{E}] \bm{D}^k\psi$ as $\psi^\dagger \epsilon^{ijk} \sigma^i \bm{D}^j [\bm{D}^l,\bm{E}^l] \bm{D}^k\psi$. The operator $g^2\psi^\dagger\bm{\sigma}\cdot\bm{B} [\bm{\partial}\cdot\bm{E}]\psi$ corresponds to two possible NRQCD operators $\frac12g^2\psi^\dagger\{\bm{\sigma}\cdot\bm{B}_aT^a, [\bm{D}^i,\bm{E}^i]_bT^b\}\psi$ and $g^2\psi^\dagger\bm{\sigma}\cdot\bm{B}_a [\bm{D}^i,\bm{E}^i]_a\psi$. The notation is such that $[\bm{D}^i,\bm{E}^i]_a=\bm{\nabla}\cdot\bm{E}_a+gf^{abc}\bm{A}_b\cdot\bm{E}_c$  \cite{Kobach:2017xkw}. Similarly $g^2\psi^\dagger[\bm{E}\cdot\bm{\partial} \bm{\sigma}\cdot\bm{B}]\psi$ corresponds to  $\frac12g^2\psi^\dagger\{\bm{E}^i_aT^a,[\bm{D}^i,\bm{\sigma}\cdot\bm{B}]_bT^b\}\psi$ and $g^2\psi^\dagger\bm{E}^i_a [\bm{D}^i,\bm{\sigma}\cdot\bm{B}]_a\psi$, $g^2\psi^\dagger[\bm{B}\cdot\bm{\partial} \bm{\sigma}\cdot\bm{E}]\psi$ corresponds to  $\frac12g^2\psi^\dagger\{\bm{B}^i_aT^a,[\bm{D}^i,\bm{\sigma}\cdot\bm{E}]_bT^b\}\psi$ and $g^2\psi^\dagger\bm{B}^i_a [\bm{D}^i,\bm{\sigma}\cdot\bm{E}]_a\psi$, $g^2\psi^\dagger[\bm{E}^i \bm{\sigma}\cdot\bm{\partial} \bm{B}^i]\psi$ corresponds to  $\frac12g^2\psi^\dagger\{\bm{E}^i_aT^a,[\bm{\sigma}\cdot\bm{D}, \bm{B}^i]_bT^b\}\psi$ and $g^2\psi^\dagger\bm{E}^i_a[\bm{\sigma}\cdot\bm{D}, \bm{B}^i]_a\psi$, and $g^2\psi^\dagger[\bm{B}^i \bm{\sigma}\cdot\bm{\partial} \bm{E}^i]\psi$ corresponds to  $\frac12g^2\psi^\dagger\{\bm{B}^i_aT^a,[\bm{\sigma}\cdot\bm{D}, \bm{E}^i]_bT^b\}\psi$ and $g^2\psi^\dagger\bm{B}^i_a[\bm{\sigma}\cdot\bm{D}, \bm{E}^i]_a\psi$. The last operator in (\ref{LNRQED4}) contains two parts: $\bm{\sigma}\cdot \bm{E}\times [{\partial_t}\bm{E}]$ and $-\bm{\sigma}\cdot \bm{E}\times[\bm{\partial}\times\bm{B} ]$. The second part can be expressed in terms of other operators in (\ref{LNRQED4}), so we will not consider it below. The first part corresponds to two possible NRQCD operators $\frac12g^2\psi^\dagger\epsilon^{ijk}\bm{\sigma}^i\bm{E}^j_a\, [{D_t},\bm{E}^k]_b\,\{T^a,T^b\}\psi$ and $g^2\psi^\dagger\epsilon^{ijk}\bm{\sigma}^i\bm{E}^j_a\, [{D_t},\bm{E}^k]_a\,\psi$. Changing $\psi\to h$, $\psi^\dagger\to \bar h$ we get 
\begin{eqnarray}\label{NRQED8SD}
&&ig \psi^\dagger  \bm{D}^i \bm{\sigma}\cdot ( \bm{D}\times\bm{E} - \bm{E}\times\bm{D} )\bm{D}^i\psi\to\nonumber\\
&&\to  \dfrac1{2M_H}(-i)\epsilon_{\rho\lambda\mu_2\mu_4}v^\rho \Pi_{\mu_1\mu_5}v_{\mu_3}\langle H |\bar h\,s^\lambda iD^{\mu_1}\{iD^{\mu_2},[iD^{\mu_3},iD^{\mu_4}]\} iD^{\mu_5}h|H\rangle=\nonumber\\
&&=12 \left(2 \tilde{a}_{12}^{(8)}- 2 \tilde{a}_{14}^{(8)}- 3 \tilde{a}_{15}^{(8)} - \tilde{b}_{13}^{(8)} + \tilde{b}_{14}^{(8)} + 3 \tilde{b}_{15}^{(8)} + \tilde{b}_{35}^{(8)} - \tilde{b}_{45}^{(8)}\right),\nonumber\\
&&ig \psi^\dagger  \epsilon^{ijk} \sigma^i \bm{D}^j [\bm{D}^l,\bm{E}^l] \bm{D}^k\psi\to\nonumber\\
&&\to  \dfrac1{2M_H}i\epsilon_{\rho\lambda\mu_1\mu_5}v^\rho \Pi_{\mu_2\mu_4}v_{\mu_3}\langle H |\bar h\,s^\lambda iD^{\mu_1}[iD^{\mu_2},[iD^{\mu_3},iD^{\mu_4}]] iD^{\mu_5}h|H\rangle=\nonumber\\
&&=12\left(2 \tilde{a}_{12}^{(8)}+ 2 \tilde{a}_{14}^{(8)}+ 3 \tilde{a}_{24}^{(8)}- \tilde{b}_{13}^{(8)}- \tilde{b}_{14}^{(8)}- 3 \tilde{b}_{34}^{(8)}- \tilde{b}_{35}^{(8)}- \tilde{b}_{45}^{(8)}\right),\nonumber
\end{eqnarray}
\begin{eqnarray}
&&\frac12g^2\psi^\dagger\{\bm{\sigma}\cdot\bm{B}_aT^a, [\bm{D}^i,\bm{E}^i]_bT^b\}\psi,\, g^2\psi^\dagger\bm{\sigma}\cdot\bm{B}_a [\bm{D}^i,\bm{E}^i]_a\psi\to\nonumber\\
&&\to  -\dfrac1{2M_H}\dfrac{i}4\epsilon_{\rho\lambda\mu_1\mu_2}v^\rho \Pi_{\mu_3\mu_5}v_{\mu_4}\langle H |\bar h\,s^\lambda \{[iD^{\mu_1},iD^{\mu_2}],[iD^{\mu_3},[iD^{\mu_4},iD^{\mu_5}]]\}h|H\rangle=\nonumber\\
&&=6\left(3 \tilde{a}_{12}^{(8)}-  \tilde{a}_{15}^{(8)}+ \tilde{a}_{24}^{(8)}-6 \tilde{b}_{13}^{(8)}- 2\tilde{b}_{14}^{(8)}+ 2\tilde{b}_{15}^{(8)}-2\tilde{b}_{34}^{(8)}+2\tilde{b}_{35}^{(8)}\right),\nonumber\\
&&\frac12g^2\psi^\dagger\{\bm{E}^i_aT^a,[\bm{D}^i,\bm{\sigma}\cdot\bm{B}]_bT^b\}\psi, \,g^2\psi^\dagger\bm{E}^i_a [\bm{D}^i,\bm{\sigma}\cdot\bm{B}]_a\psi\to\nonumber\\
&&\to  -\dfrac1{2M_H}\dfrac{i}4\epsilon_{\rho\lambda\mu_4\mu_5}v^\rho \Pi_{\mu_2\mu_3}v_{\mu_1}\langle H |\bar h\,s^\lambda \{[iD^{\mu_1},iD^{\mu_2}],[iD^{\mu_3},[iD^{\mu_4},iD^{\mu_5}]]\}h|H\rangle=\nonumber\\
&&=6\left(4 \tilde{b}_{13}^{(8)}-4\tilde{b}_{15}^{(8)}+\tilde{b}_{34}^{(8)}-2\tilde{b}_{35}+\tilde{b}_{45}\right),\nonumber\\
&&\frac12g^2\psi^\dagger\{\bm{B}^i_aT^a,[\bm{D}^i,\bm{\sigma}\cdot\bm{E}]_bT^b\}\psi, \,g^2\psi^\dagger\bm{B}^i_a [\bm{D}^i,\bm{\sigma}\cdot\bm{E}]_a\psi\to\nonumber\\
&&\to  \dfrac1{2M_H}\dfrac{i}4\epsilon_{\rho\mu_1\mu_2\mu_3}v^\rho \Pi_{\lambda\mu_5}v_{\mu_4}\langle H |\bar h\,s^\lambda \{[iD^{\mu_1},iD^{\mu_2}],[iD^{\mu_3},[iD^{\mu_4},iD^{\mu_5}]]\}h|H\rangle=\nonumber\\
&&=-6\left(\tilde{a}_{12}^{(8)}+ \tilde{a}_{14}^{(8)}- \tilde{a}_{24}^{(8)}-2 \tilde{b}_{13}^{(8)}+\tilde{b}_{14}^{(8)}-\tilde{b}_{15}^{(8)}+\tilde{b}_{34}^{(8)}-\tilde{b}_{35}^{(8)}\right),\nonumber\\
&&\frac12g^2\psi^\dagger\{\bm{E}^i_aT^a,[\bm{\sigma}\cdot\bm{D}, \bm{B}^i]_bT^b\}\psi,\, g^2\psi^\dagger\bm{E}^i_a[\bm{\sigma}\cdot\bm{D}, \bm{B}^i]_a\psi\to\nonumber\\
&&\to  \dfrac1{2M_H}\dfrac{i}4\epsilon_{\rho\mu_2\mu_4\mu_5}v^\rho \Pi_{\lambda\mu_3}v_{\mu_1}\langle H |\bar h\,s^\lambda \{[iD^{\mu_1},iD^{\mu_2}],[iD^{\mu_3},[iD^{\mu_4},iD^{\mu_5}]]\}h|H\rangle=\nonumber\\
&&=-6\left(\tilde{b}_{13}^{(8)}-\tilde{b}_{15}^{(8)}-\tilde{b}_{34}^{(8)}+2\tilde{b}_{35}^{(8)}-\tilde{b}_{45}^{(8)}\right),\nonumber\\
&&\frac12g^2\psi^\dagger\{\bm{B}^i_aT^a,[\bm{\sigma}\cdot\bm{D}, \bm{E}^i]_bT^b\}\psi,\,g^2\psi^\dagger\bm{B}^i_a[\bm{\sigma}\cdot\bm{D}, \bm{E}^i]_a\psi\to\nonumber\\
&&\to  \dfrac1{2M_H}\dfrac{i}4\epsilon_{\rho\mu_1\mu_2\mu_5}v^\rho \Pi_{\lambda\mu_3}v_{\mu_4}\langle H |\bar h\,s^\lambda \{[iD^{\mu_1},iD^{\mu_2}],[iD^{\mu_3},[iD^{\mu_4},iD^{\mu_5}]]\}h|H\rangle=\nonumber\\
&&=-6\left(\tilde{a}_{12}^{(8)}-\tilde{a}_{14}^{(8)}+ \tilde{a}_{15}^{(8)}-2\tilde{b}_{13}^{(8)}+\tilde{b}_{14}^{(8)}-\tilde{b}_{15}^{(8)}+\tilde{b}_{34}^{(8)}-\tilde{b}_{35}^{(8)}\right),\nonumber\\
&&\frac12g^2\psi^\dagger\epsilon^{ijk}\bm{\sigma}^i\bm{E}^j_a\, [{D_t},\bm{E}^k]_b\,\{T^a,T^b\}\psi,\,g^2\psi^\dagger\epsilon^{ijk}\bm{\sigma}^i\bm{E}^j_a\, [{D_t},\bm{E}^k]_a\,\psi \to\nonumber\\
&&\to  -\dfrac1{2M_H}\dfrac{i}2\epsilon_{\rho\lambda\mu_2\mu_5}v^\rho v_{\mu_1}v_{\mu_3}v_{\mu_4}\langle H |\bar h\,s^\lambda \{[iD^{\mu_1},iD^{\mu_2}],[iD^{\mu_3},[iD^{\mu_4},iD^{\mu_5}]]\}h|H\rangle=\nonumber\\
&&=6\tilde{c}^{(8)}.
\end{eqnarray}
As in the previous cases, one can check and verify that the matrix elements of operators that have two color structures depend on linear combinations of parameters calculated in  \ref{sec_SD8}. This confirms our observation that one can predict how many operators have multiple color structures using our general method. 

We anticipated seventeen operators in section \ref{sec_SD8}, but generalizing the NRQED to the NRQCD case gives only fourteen operators. It is clear the NRQCD Lagrangian will contain three extra operators.  We will list them in section \ref{NRQCD_New}.

\section{Applications}\label{Applications}
\subsection{Dimension nine spin independent HQET operators} \label{SI9}
Based on the general method, it is easy to parameterize the matrix element of the general spin-independent dimension nine HQET operators. We have six covariant derivatives and we can have terms with zero $v$'s, two $v$'s, or 4 $v$'s. Taking into account the various ways to contract the other indices, we have all together 24 possible tensors: \begin{eqnarray} \label{dim9}
&&\dfrac1{2M_H}\langle H |\bar h\, iD^{\mu_1}iD^{\mu_2}iD^{\mu_3}iD^{\mu_4}iD^{\mu_5}iD^{\mu_6}h|H\rangle=a^{(9)}_{12,34}\,\Pi^{\mu_1\mu_2}\Pi^{\mu_3\mu_4}\Pi^{\mu_5\mu_6}+\nonumber\\
&&+ a^{(9)}_{12,35}\,
\left(\Pi^{\mu_1\mu_2}\Pi^{\mu_3\mu_5}\Pi^{\mu_4\mu_6}+\Pi^{\mu_1\mu_3}\Pi^{\mu_2\mu_4}\Pi^{\mu_5\mu_6}\right)+a^{(9)}_{12,36}\,\left(\Pi^{\mu_1\mu_2}\Pi^{\mu_3\mu_6}\Pi^{\mu_4\mu_5}+\Pi^{\mu_1\mu_4}\Pi^{\mu_2\mu_3}\Pi^{\mu_5\mu_6}\right)+\nonumber\\
&&+ a^{(9)}_{13,25}\,\Pi^{\mu_1\mu_3}\Pi^{\mu_2\mu_5}\Pi^{\mu_4\mu_6}+a^{(9)}_{13,26}\,\left(\Pi^{\mu_1\mu_3}\Pi^{\mu_2\mu_6}\Pi^{\mu_4\mu_5}+\Pi^{\mu_1\mu_5}\Pi^{\mu_2\mu_3}\Pi^{\mu_4\mu_6}\right)+\nonumber\\
&&+a^{(9)}_{14,25}\,\Pi^{\mu_1\mu_4}\Pi^{\mu_2\mu_5}\Pi^{\mu_3\mu_6}+a^{(9)}_{14,26}\,
\left(\Pi^{\mu_1\mu_4}\Pi^{\mu_2\mu_6}\Pi^{\mu_3\mu_5}+\Pi^{\mu_1\mu_5}\Pi^{\mu_2\mu_4}\Pi^{\mu_3\mu_6}\right)+\nonumber\\
&&+a^{(9)}_{15,26}\,\Pi^{\mu_1\mu_5}\Pi^{\mu_2\mu_6}\Pi^{\mu_3\mu_4}+a^{(9)}_{16,23}\,\Pi^{\mu_1\mu_6}\Pi^{\mu_2\mu_3}\Pi^{\mu_4\mu_5}+a^{(9)}_{16,24}\,\Pi^{\mu_1\mu_6}\Pi^{\mu_2\mu_4}\Pi^{\mu_3\mu_5}+\nonumber\\
&&+a^{(9)}_{16,25}\,\Pi^{\mu_1\mu_6}\Pi^{\mu_2\mu_5}\Pi^{\mu_3\mu_4}+b^{(9)}_{12,36}\,\left(\Pi^{\mu_1\mu_2}\Pi^{\mu_3\mu_6}v^{\mu_4}v^{\mu_5}+\Pi^{\mu_1\mu_4}\Pi^{\mu_5\mu_6}v^{\mu_2}v^{\mu_3}\right)+\nonumber\\
&&+b^{(9)}_{12,46}\,\left(\Pi^{\mu_1\mu_2}\Pi^{\mu_4\mu_6}v^{\mu_3}v^{\mu_5}+\Pi^{\mu_1\mu_3}\Pi^{\mu_5\mu_6}v^{\mu_2}v^{\mu_4}\right)+b^{(9)}_{12,56}\,\Pi^{\mu_1\mu_2}\Pi^{\mu_5\mu_6}v^{\mu_3}v^{\mu_4}+\nonumber\\
&&+b^{(9)}_{13,26}\,\left(\Pi^{\mu_1\mu_3}\Pi^{\mu_2\mu_6}v^{\mu_4}v^{\mu_5}+\Pi^{\mu_1\mu_5}\Pi^{\mu_4\mu_6}v^{\mu_2}v^{\mu_3}\right)+\nonumber\\
&&+b^{(9)}_{13,46}\,\Pi^{\mu_1\mu_3}\Pi^{\mu_4\mu_6}v^{\mu_2}v^{\mu_5}+b^{(9)}_{14,26}\,\left(\Pi^{\mu_1\mu_4}\Pi^{\mu_2\mu_6}v^{\mu_3}v^{\mu_5}+\Pi^{\mu_1\mu_5}\Pi^{\mu_3\mu_6}v^{\mu_2}v^{\mu_4}\right)+\nonumber\\
&&b^{(9)}_{14,36}\,\Pi^{\mu_1\mu_4}\Pi^{\mu_3\mu_6}v^{\mu_2}v^{\mu_5}+b^{(9)}_{15,26}\,\Pi^{\mu_1\mu_5}\Pi^{\mu_2\mu_6}v^{\mu_3}v^{\mu_4}+\nonumber\\
&&b^{(9)}_{16,23}\,\left(\Pi^{\mu_1\mu_6}\Pi^{\mu_2\mu_3}v^{\mu_4}v^{\mu_5}+\Pi^{\mu_1\mu_6}\Pi^{\mu_4\mu_5}v^{\mu_2}v^{\mu_3}\right)+\nonumber\\
&&+b^{(9)}_{16,24}\,\left(\Pi^{\mu_1\mu_6}\Pi^{\mu_2\mu_4}v^{\mu_3}v^{\mu_5}+\Pi^{\mu_1\mu_6}\Pi^{\mu_3\mu_5}v^{\mu_2}v^{\mu_4}\right)+b^{(9)}_{16,25}\,\Pi^{\mu_1\mu_6}\Pi^{\mu_2\mu_5}v^{\mu_3}v^{\mu_4}+\nonumber\\
&&+b^{(9)}_{16,34}\,\Pi^{\mu_1\mu_6}\Pi^{\mu_3\mu_4}v^{\mu_2}v^{\mu_5}+c^{(9)}\,\Pi^{\mu_1\mu_6}v^{\mu_2}v^{\mu_3}v^{\mu_4}v^{\mu_5}.
\end{eqnarray}
Our notation is such that the subscripts denotes the first indices that are contracted via $\Pi$'s in numerical order. We use different letters for different number of $v$'s in the tensors.

We should in principle consider also the various possible color structures. These can arise from combining three possible pure color octets:  $[iD^{\mu_i},iD^{\mu_j}]$, $[iD^{\mu_i},[iD^{\mu_j},iD^{\mu_k}]]$, and $[iD^{\mu_i},[iD^{\mu_j},[iD^{\mu_k},iD^{\mu_l}]]]$. There are several possibilities for combining them with each other and/or with other covariant derivatives. We will not consider these multiple color structure here. For phenomenological applications at the current level of precision, see e.g. section \ref{sec:moments}, only one color structure is needed, namely, the one that contains $T^aT^b$ and not $\delta^{ab}$.

\subsection{Moments of the leading power shape functions}\label{sec:moments}
In analyzing charmless inclusive $B$ decays one often encounters ``shape functions" \cite{Neubert:1993ch,Neubert:1993um,Bigi:1993ex,Bauer:2001mh,Lee:2004ja,Bosch:2004cb, Beneke:2004in,Benzke:2010js}. These are Fourier transforms of diagonal matrix elements of non-local HQET operators, analogous to nucleon parton distribution functions. Moments of these shape functions can often be related to HQET parameters. The matrix elements decomposition presented above makes the calculation of the moments especially easy. We illustrate this by calculating moments of of the leading power shape. 

The leading power shape function can be defined as\footnote{There are several equivalent definitions of the leading power shape functions in the literature. The one presented here can be obtained from \cite{Benzke:2010js} by using the translation invariance of the matrix element elements and changing $t\to-t$ in the integration.}
\begin{equation}
   S(\omega)
   = \int\frac{dt}{2\pi}\,e^{i\omega t}\,
   \frac{\langle\bar B(v)| \bar h(0) S_n(0) S_n^\dagger(tn) h(tn)
   |\bar B(v)\rangle }{2M_B} \,,
\end{equation}
where is  $n$ is a light-like vector, i.e. $n^2=0$, and $n\cdot v=1$. $S_n$ is a Wilson line in the $n$ direction, see  \cite{Benzke:2010js} for its definition.  

For completeness we review how the moments of $S(\omega)$ are related to local HQET operators.  The zeroth moment of $S(\omega)$ is 
\begin{eqnarray}
   \int d\omega\, S(\omega)
   &=& \int d\omega \int\frac{dt}{2\pi}\,e^{i\omega t}\,
   \frac{\langle\bar B(v)| \bar h(0) S_n(0) S_n^\dagger(tn) h(tn)
   |\bar B(v)\rangle }{2M_B}= \nonumber\\&=&
   \frac{\langle\bar B(v)| \bar h(0) S_n(0) S_n^\dagger(0) h(0)
   |\bar B(v)\rangle }{2M_B} =\frac{\langle\bar B(v)| \bar h(0) h(0)
   |\bar B(v)\rangle }{2M_B},
\end{eqnarray}
where we have used the unitarity of the Wilson lines,  $S_n(x) S_n^\dagger(x)=1$. Using the identity, $in\cdot D\,S_n(x)= S_n(x)\,in\cdot \partial$, the first moment of $S(\omega)$ is 
\begin{eqnarray}
   \int d\omega\, \omega\, S(\omega)
 &=& \int dt \int\frac{d\omega}{2\pi}\,\left(-i\dfrac{\partial}{\partial t}e^{i\omega t}\right)\,
   \frac{\langle\bar B(v)| \bar h(0) S_n(0)S_n^\dagger(tn) h(tn)
   |\bar B(v)\rangle }{2M_B}= \nonumber\\
   &=& \int dt \int\frac{d\omega}{2\pi}\,e^{i\omega t}\,
   \frac{\langle\bar B(v)| \bar h(0) S_n(0)\, i n\cdot \partial\, S_n^\dagger(tn) h(tn)
   |\bar B(v)\rangle }{2M_B}=\nonumber\\
   &=& \int dt \int\frac{d\omega}{2\pi}\,e^{i\omega t}\,
   \frac{\langle\bar B(v)| \bar h(0) S_n(0)S_n^\dagger(tn)S_n(tn)\, i n\cdot \partial\, S_n^\dagger(tn) h(tn)
   |\bar B(v)\rangle }{2M_B}=\nonumber\\
    &=& \int dt \int\frac{d\omega}{2\pi}\,e^{i\omega t}\,
   \frac{\langle\bar B(v)| \bar h(0) S_n(0)S_n^\dagger(tn)\,i n\cdot D\, S_n(tn)S_n^\dagger(tn) h(tn)
   |\bar B(v)\rangle }{2M_B}=\nonumber\\
    &=&  \frac{\langle\bar B(v)| \bar h(0) \,i n\cdot D\, h(0)
   |\bar B(v)\rangle }{2M_B}.
   \end{eqnarray}
Similarly we can show that the $k$-th moment of the $S(\omega)$ is 
\begin{equation}\label{general_moment}
 \int d\omega\, \omega^k\, S(\omega)= \frac{\langle\bar B(v)| \bar h\,( i n\cdot D)^k\, 
  h |\bar B(v)\rangle }{2M_B}=\frac{n_{\mu_1}...n_{\mu_k}\langle\bar B(v)| \bar h \,i D^{\mu_1}...i D^{\mu_k}\, h
   |\bar B(v)\rangle }{2M_B}.
\end{equation}
Using (\ref{general_moment})  we can easily calculate the moments. We find that the first six moments are
\begin{eqnarray}\label{moments}
\int d\omega\,  S(\omega)&=&1,\qquad \int d\omega\, \omega\, S(\omega)=0,\qquad \int d\omega\, \omega^2\, S(\omega)=-a^{(5)}=-\lambda_1/3,\nonumber\\
\int d\omega\, \omega^3\, S(\omega)&=&-a^{(6)}=-\rho_1/3,\qquad 
\int d\omega\, \omega^4\, S(\omega)=a_{12}^{(7)} + a_{13}^{(7)} + a_{14}^{(7)} -b^{(7)}=m_1/5-m_2/3,\nonumber\\
\int d\omega\, \omega^5\, S(\omega)&=&2a_{12}^{(8)} + 2a_{13}^{(8)} + 2a_{15}^{(8)}+b_{12}^{(8)} + b_{14}^{(8)} + b_{15}^{(8)}-c^{(8)}=\nonumber\\
&=&\left(-8 r_1 + 2 r_2 + 2 r_3 + 2 r_4 + r_5 + r_6 + r_7\right)/15,\nonumber\\
\int d\omega\, \omega^6\, S(\omega)&=&-a^{(9)}_{12,34} - 2 a^{(9)}_{12,35} - 2 a^{(9)}_{12,36} - a^{(9)}_{13,25} - 2a^{(9)}_{13,26} - a^{(9)}_{14,25} - 2 a^{(9)}_{14,26} - a^{(9)}_{15,26} - a^{(9)}_{16,23}\nonumber\\
 && - a^{(9)}_{16,24} - a^{(9)}_{16,25} +2b^{(9)}_ {12,36} +
 2 b^{(9)}_{12,46} +b^{(9)}_{12,56} + 2b^{(9)}_{13,26} +b^{(9)}_{13,46} +2b^{(9)}_{14,26} + b^{(9)}_{14,36}  \nonumber\\
 &&+b^{(9)}_{15,26}+ 2 b^{(9)}_{16,23} +2 b^{(9)}_{16,24} + b^{(9)}_{16,25} +b^{(9)}_{16,34} - c^{(9)}.
\end{eqnarray}
For all the moments apart from the sixth we have used the relations  to previously defined HQET parameters from section \ref{HQET}. The sixth moment is expressed in terms of the new HQET parameters of section \ref{SI9}. Using the HQET parameters extracted in \cite{Gambino:2016jkc} one can use the moments up to the fifth one to improve the modeling of the leading power shape function.  

\subsection{NRQCD Lagrangian to order $1/M^4$}\label{NRQCD_New}
From section \ref{NRQED} we learn that the $1/M^4$ NRQCD Lagrangian contains three spin-independent operators and three spin-dependent operators that cannot be obtained from simple generalization of the  NRQED Lagrangian. Here we list these new operators.

The new NRQCD operators contain commutators of chromoelectric and chromomagnetic fields, which vanish for NRQED.  We can easily test the linear independence of the possible operators  by calculating the matrix elements of the corresponding HQET operators. For the spin-independent operators we have 
\begin{eqnarray}\label{NRQCD8SI}
&&g^2 \psi^\dagger  [\bm{E}^i,[iD_t,\bm{E}^i] ]_aT^a\psi\to\nonumber\\
&&\to\dfrac1{2M_H}\langle H |\bar h\,  [[iD^{\mu_1},iD^{\mu_2}], [iD^{\mu_3},[iD^{\mu_4},iD^{\mu_5}]]]h|H\rangle v_{\mu_1}v_{\mu_3}v_{\mu_4}\Pi_{\mu_2\mu_5}=-6c^{(8)},\nonumber\\
&&ig^2\psi^\dagger [\bm{B}^i,( \bm{D} \times \bm{E} + \bm{E}\times \bm{D} )^i]_aT^a \psi\to \nonumber\\
&&\to\dfrac1{2M_H}\langle H |\bar h\,  [[iD^{\mu_1},iD^{\mu_2}], [iD^{\mu_3},[iD^{\mu_4},iD^{\mu_5}]]]h|H\rangle v_{\mu_4}\Pi_{\mu_1\mu_3}\Pi_{\mu_2\mu_5}=12(b_{14}^{(8)}-b_{15}^{(8)}),\nonumber\\
&&ig^2\psi^\dagger [\bm{E}^i,( \bm{D} \times \bm{B} + \bm{B}\times \bm{D} )^i]_aT^a \psi\to\nonumber\\
&&\to -\dfrac1{2M_H}\langle H |\bar h\,  [[iD^{\mu_1},iD^{\mu_2}], [iD^{\mu_3},[iD^{\mu_4},iD^{\mu_5}]]]h|H\rangle v_{\mu_1}\Pi_{\mu_3\mu_4}\Pi_{\mu_2\mu_5}=\nonumber\\
&&=12(a_{12}^{(8)}-2a_{13}^{(8)}+a_{15}^{(8)}).
\end{eqnarray}
It is easy to check that these operators are linearly independent of the operators of (\ref{NRQED8SI}) and  Hermitian and invariant under $P$ and $T$. 

For the spin-dependent operators we consider the set of spin-dependent \emph{NRQED} operators $O_{X7}\equiv\dfrac{1}{2}g^2 \psi^\dagger \{\bm{\sigma}\cdot\bm{B}, [\bm{D}^i,\bm{E}^i]\}\psi$, $O_{X8}\equiv\dfrac{1}{2}g^2  \psi^\dagger\{\bm{E}^i,[\bm{D}^i,\bm{\sigma}\cdot\bm{B}]\}\psi$, $O_{X9}\equiv\dfrac{1}{2}g^2  \psi^\dagger \{\bm{B}^i,[\bm{D}^i,\bm{\sigma}\cdot\bm{E}]\}\psi$, $O_{X10}\equiv\dfrac{1}{2}g^2  \psi^\dagger \{\bm{E}^i, [\bm{\sigma}\cdot\bm{D}, \bm{B}^i]\}\psi$, $O_{X11}\equiv\dfrac{1}{2}g^2  \psi^\dagger \{\bm{B}^i, [\bm{\sigma}\cdot\bm{D}, \bm{E}^i]\}\psi$, where the notations follows from equation (\ref{LNRQED4}). We can generate new operators by  replacing the commutators by anti-commutators and vice versa.  Only three operators will be linearly independent of the NRQED spin-dependent operators.  We can choose to modify $O_{X7}, O_{X9}, O_{X10}$, or $O_{X7}, O_{X9}, O_{X11}$, or $O_{X8}, O_{X9}, O_{X10}$, or $O_{X8}, O_{X9}, O_{X11}$. We choose $O_{X7}, O_{X9}, O_{X10}$. We have 
\begin{eqnarray}\label{NRQCD8SD}
&&g^2 \psi^\dagger [\bm{\sigma}\cdot\bm{B}, \{\bm{D}^i,\bm{E}^i\}]_aT^a\psi\to\nonumber\\
&&\to  -\dfrac1{2M_H}\dfrac{i}2\epsilon_{\rho\lambda\mu_1\mu_2}v^\rho \Pi_{\mu_3\mu_5}v_{\mu_4}\langle H |\bar h\,s^\lambda [[iD^{\mu_1},iD^{\mu_2}],\{iD^{\mu_3},[iD^{\mu_4},iD^{\mu_5}]\}]h|H\rangle=\nonumber\\
&&=-12\left(3 \tilde{a}_{12}^{(8)}- \tilde{a}_{15}+ \tilde{a}_{24}\right),\nonumber\\
&&g^2  \psi^\dagger [\bm{B}^i,\{\bm{D}^i,\bm{\sigma}\cdot\bm{E}\}]_aT^a\psi\to\nonumber\\
&&\to  \dfrac1{2M_H}\dfrac{i}2\epsilon_{\rho\mu_1\mu_2\mu_3}v^\rho \Pi_{\lambda\mu_5}v_{\mu_4}\langle H |\bar h\,s^\lambda [[iD^{\mu_1},iD^{\mu_2}],\{iD^{\mu_3},[iD^{\mu_4},iD^{\mu_5}]\}]h|H\rangle=\nonumber\\
&&=12\left(\tilde{a}_{12}^{(8)}+ \tilde{a}_{14}^{(8)}- \tilde{a}_{24}-\tilde{b}_{14}+\tilde{b}_{15}+\tilde{b}_{34}-\tilde{b}_{35}\right),\nonumber\\
&&g^2  \psi^\dagger [\bm{E}^i, \{\bm{\sigma}\cdot\bm{D}, \bm{B}^i\}]_aT^a\psi\to\nonumber\\
&&\to  \dfrac1{2M_H}\dfrac{i}2\epsilon_{\rho\mu_2\mu_4\mu_5}v^\rho \Pi_{\lambda\mu_3}v_{\mu_1}\langle H |\bar h\,s^\lambda [iD^{\mu_1},iD^{\mu_2}],\{iD^{\mu_3},[iD^{\mu_4},iD^{\mu_5}]\}]h|H\rangle=\nonumber\\
&&=-12\left(\tilde{b}_{13}+\tilde{b}_{15}-\tilde{b}_{34}+\tilde{b}_{45}\right).
\end{eqnarray}
 It is easy to check that these operators are linearly independent of the operators of (\ref{NRQED8SD}) and Hermitian and invariant under $P$ and $T$. 
  
 We can now list the full dimension-eight  NRQCD Lagrangian. It is
\begin{align}\label{LNRQCD4}
&{\cal L}_{\mbox{\scriptsize NRQCD}}^{\mbox{\scriptsize dim$=$8}} = \psi^\dagger
  \bigg\{ 
c_{X1}g { [ \bm{D}^2 , \{\bm{D}^i, \bm{E}^i\}] \over M^4 }
+ c_{X2}g {  \{ \bm{D}^2 , [\bm{D}^i, \bm{E}^i]\} \over M^4 }+ c_{X3}g {[\bm{D}^i, [\bm{D}^i, [\bm{D}^j,\bm{E}^j]] ]\over M^4 }
\nonumber\\
&\quad 
+ i c_{X4a}\,g^2 {  \{ \bm{D}^i , \epsilon^{ijk}\bm{E}_a^j\bm{B}_b^k\,\{T^a,T^b\}\}\over 2M^4 } + i c_{X4b}\,g^2 {\{ \bm{D}^i , \epsilon^{ijk}\bm{E}_a^j\bm{B}_b^k\,\delta^{ab}\} \over M^4 } 
\nonumber\\
&\quad
+ ic_{X5} g { \bm{D}^i \bm{\sigma}\cdot ( \bm{D}\times\bm{E} - \bm{E}\times\bm{D} )\bm{D}^i   \over M^4} + ic_{X6} g { \epsilon^{ijk} \sigma^i \bm{D}^j [\bm{D}^l,\bm{E}^l] \bm{D}^k \over M^4} 
\nonumber\\
&\quad
+c_{X7a}\, g^2 { \{\bm{\sigma}\cdot\bm{B}_aT^a, [\bm{D}^i,\bm{E}^i]_bT^b\} \over 2M^4}+c_{X7b}\,  g^2 { \bm{\sigma}\cdot\bm{B}_a [\bm{D}^i,\bm{E}^i]_a \over M^4} 
 \nonumber\\
&\quad
+  c_{X8a}\, g^2 { \{\bm{E}^i_aT^a,[\bm{D}^i,\bm{\sigma}\cdot\bm{B}]_bT^b\}\over 2M^4}+  c_{X8b} g^2 {\bm{E}^i_a [\bm{D}^i,\bm{\sigma}\cdot\bm{B}]_a\over M^4}
\nonumber\\
&\quad
+ c_{X9a}\, g^2 { \{\bm{B}^i_aT^a,[\bm{D}^i,\bm{\sigma}\cdot\bm{E}]_bT^b\}\over 2M^4} 
+ c_{X9b}\, g^2 {\bm{B}^i_a [\bm{D}^i,\bm{\sigma}\cdot\bm{E}]_a\over M^4} 
\nonumber\\
&\quad
+ c_{X10a}\, g^2 {\{\bm{E}^i_aT^a,[\bm{\sigma}\cdot\bm{D}, \bm{B}^i]_bT^b\}\over 2M^4} 
+ c_{X10b}\, g^2 {\bm{E}^i_a[\bm{\sigma}\cdot\bm{D}, \bm{B}^i]_a\over M^4} 
\nonumber\\
&\quad
+ c_{X11a}\, g^2 {\{\bm{B}^i_aT^a,[\bm{\sigma}\cdot\bm{D}, \bm{E}^i]_bT^b\}\over 2M^4} 
+ c_{X11b}\, g^2 {\bm{B}^i_a[\bm{\sigma}\cdot\bm{D}, \bm{E}^i]_a\over M^4} 
 \nonumber\\
&\quad
+ \tilde{c}_{X12a}\, g^2 { \epsilon^{ijk}\bm{\sigma}^i\bm{E}^j_a\, [{D_t},\bm{E}^k]_b\,\{T^a,T^b\}\over 2M^4}
+ \tilde{c}_{X12b}\, g^2 { \epsilon^{ijk}\bm{\sigma}^i\bm{E}^j_a\, [{D_t},\bm{E}^k]_a\over M^4}  \nonumber\\
&\quad
+i c_{X13} g^2 {  [\bm{E}^i,[D_t,\bm{E}^i] ]\ \over M^4}
+ ic_{X14} g^2 { [\bm{B}^i,( \bm{D} \times \bm{E} + \bm{E}\times \bm{D} )^i] \over M^4} 
+ ic_{X15} g^2 {  [\bm{E}^i,( \bm{D} \times \bm{B} + \bm{B}\times \bm{D} )^i] \over M^4} \nonumber\\
&\quad
+ c_{X16} g^2 { [\bm{\sigma}\cdot\bm{B}, \{\bm{D}^i,\bm{E}^i\}]\over M^4}
+ c_{X17} g^2 { [\bm{B}^i,\{\bm{D}^i,\bm{\sigma}\cdot\bm{E}\}] \over M^4} 
+ c_{X18} g^2 { [\bm{E}^i, \{\bm{\sigma}\cdot\bm{D}, \bm{B}^i\}] \over M^4} \bigg\} \psi.
\end{align}
Notice that we have modified the operators $O_{X12}$ as explained in section \ref{NRQCD_SD8}. We have introduced extra $a$ and $b$ subscripts for operators that have multiple color structure but the same Lorentz structure.  Explicit color indices are shown for these operators. These results agree also with \cite{Kobach:2017xkw} that uses a slightly different basis.

\section{Conclusions} \label{Conclusions}
Effective field theories are an important tool in current research. The effective theory Lagrangian is often written as a series of operators with increasing dimension suppressed by the inverse powers of the cutoff scale of the theory. As research progresses, higher dimensional operators are receiving more and more attention. For example, the construction of higher dimension operators in the Standard Model Effective Field Theory (SM EFT) was shown recently be simpler than one might expect \cite{Henning:2015alf}.

In this paper we investigated the question of constructing higher dimensional operators for the HQET and NRQCD (NRQED) Lagrangians. Despite having a different power counting, the two Lagrangians are closely related. We showed how one can analyze operators that contain two HQET fields or two NRQCD (NRQED) fields with an arbitrary number of covariant derivatives. The method we use is to consider diagonal matrix elements of HQET operators between pseudo-scalar meson states. We write such matrix elements as non-perturbative HQET parameters multiplied by tensors constructed from the heavy quark velocity, the metric tensor and the Levi-Civita tensor. Imposing constraints from $P$ and $T$ symmetries, hermitian conjugation, and the fact that we consider theories in 3+1 dimensions, allows the reduce the number of HQET parameters of a given dimension. The number of possible HQET operators at each dimension corresponds to the number of HQET parameters. This method allows us to easily determine the number of operators at each dimension and whether a given set of operators of a given dimension is linearly independent.  NRQCD and NRQED operators can be similarly analyzed by replacing those fields with HQET fields and considering the matrix elements of these operators.

One drawback of this method is that it does not distinguish operators that have the same Lorentz structure but different color structure. As was recently pointed out in \cite{Kobach:2017xkw}, operators that contain a symmetric product of two color matrices, e.g.  $\psi^\dagger E^i_aT^a  E^i_bT^b\psi$, can be decomposed in terms of a color octet and a color singlet operators, e.g.   $\psi^\dagger E^i_a E^i_b\,d^{abc}T^c \psi$ and $\psi^\dagger E^i_a E^i_b \delta^{ab}\psi$. Since they only differ in their color structure, both will give the same linear combination of parameters.  Alternatively we can use the basis of  $\psi^\dagger E^i_a E^i_b\left\{T^a,T^b\right\}\psi$ and $\psi^\dagger E^i_a E^i_b \delta^{ab}\psi$. The operator $\psi^\dagger E^i_a E^i_b\left\{T^a,T^b\right\} \psi$ is generated by commutator and anti-commutators of covariant derivatives and it is the only of the two that appears when calculating observables at tree level.  The operator $\psi^\dagger E^i_a E^i_b \delta^{ab}\psi$ will be generated when considering radiative corrections \cite{Manohar:1997qy}. For applications to inclusive $B$ decays this operator arises at only ${\cal O}(\alpha_s)/m_b^4$, beyond the current level of precision. This explains why such operators were not considered in \cite{Mannel:2010wj}. To address the possibility of multiple color structures, one has to consider them separately from the general method we presented. But using the method presented above allows to determine how many linearly independent operators there are for possible different color structures.  We showed how this is done for the dimension seven and eight operators and confirmed the results of \cite{Kobach:2017xkw} for the multiple color structures.  

We presented our general method in section \ref{General}. We demonstrated it by relating the HQET parameters of operators of dimension four, five, six, seven, and eight known from the literature to our basis in section \ref{HQET}. NRQCD operators up to dimension seven and NRQED operators up to dimension eight were previously known in the literature. We have analyzed these operators and related the corresponding HQET matrix elements to our basis in section \ref{NRQED}. This allows to easily relate the known HQET and NRQCD operators to one another.

Going beyond the known operators, we presented several new results in section \ref{Applications}.  We analyzed the dimension nine spin-independent HQET parameters, finding 24 possible parameters (not including multiple color structures) see equation (\ref{dim9}). We calculated moments of the leading power shape function in terms of HQET parameters up to and including local matrix elements of dimension nine, see equation (\ref{moments}). This will allow to improve the modeling of the leading power shape function. Similarly, one can use this to improve modeling of subleading shape functions \cite{in_progress}. Most importantly, we constructed the dimension eight NRQCD operators that do not appear in the $1/M^4$ NRQED Lagrangian. These allow to present for the full $1/M^4$ bilinear NRQCD Lagrangian, see equation (\ref{LNRQCD4}).

We conclude by considering possible extensions of this work. The method we presented allows in principle to write down all the possible HQET operators of any given dimension. It would be interesting to automatize the procedure using a computer program to construct these higher dimensional operators and of the NRQCD Lagrangian. Also, certain multiple color structures were considered separately from the general method. It would be desirable to find a method that automatically generate these color structures.

A separate interesting question is what are the Wilson coefficients of the operators. In particular what are the relations between coefficients of operators of different dimensions. These are known as ``reparameterization invariance"  \cite{Luke:1992cs} or ``Lorentz invariance" constraints\footnote{The ansatz for implementing Lorentz invariance via reparametrization invariance breaks down starting at dimension eight operators, see  \cite{Heinonen:2012km}.}  \cite{Heinonen:2012km}. For NRQED such relations are known for up to dimension eight operators \cite{Heinonen:2012km, Hill:2012rh} but not for NRQCD or HQET operators above dimension six.  Such relations allow to determine the contribution of a certain higher dimensional operators based on the knowledge of lower dimensional operators. This has applications to semileptonic and radiative $B$ decays, see e.g. \cite{Becher:2007tk, Ewerth:2009yr, Manohar:2010sf}.  

 Finally, throughout  the paper we have not considered operators with more than two HQET or NRQCD (NRQED) fields. The reason, as explained in the introduction, is that the one non-relativistic fermion sector can be combined with an additional non-relativistic field or an additional relativistic field.  Results for each case were presented in the literature  \cite{Bodwin:1994jh, Brambilla:2006ph, Brambilla:2008zg, Hill:2012rh, Dye:2016uep}, but not for an arbitrary operator dimension.

\vskip 0.2in
\noindent
{\bf Acknowledgements}
\vskip 0.1in
\noindent
We  thank Andrew E. Blechman and Alexey A. Petrov for useful discussions and comments on the manuscript. We also thank the authors of \cite{Kobach:2017xkw} for pointing out the missing operators in the first version of this paper. This work was supported by DOE grant DE-SC0007983.

\end{document}